%% file: main.tex
\documentclass[%
 reprint,
 amsmath,amssymb,
 aps,
]{revtex4-1}

\usepackage{graphicx}% Include figure files
\usepackage{dcolumn}% Align table columns on decimal point
\usepackage{bm}% bold math
\usepackage{amsmath}
\usepackage{subcaption}
\usepackage{float}
\usepackage[justification=justified,
   format=plain]{caption}
\begin{document}

\preprint{APS/123-QED}

\title{Sideband cooling of the radial modes of motion of a single ion in a Penning trap}% Force line breaks with \\
%\thanks{A footnote to the article title}

\author{P. Hrmo} 
\email{Now at: Institute of Experimental Physics, Universit of Innsbruck, Innsbruck, Austria}
\author{M. K. Joshi}
\email{Now at: Institute for Quantum Optics and Quantum Information, Austrian Academy of Sciences, Innsbruck, Austria}%
\author{V. Jarlaud}%
\email{Now at: Aarhus University, Aarhus, Denmark}
\author{O. Corfield}%
\author{R. C. Thompson}
\email{r.thompson@imperial.ac.uk}%
\affiliation{Quantum Optics and Laser Science, Blackett Laboratory,\\
Imperial College London, Prince Consort Road, London, SW7 2AZ, United Kingdom
}%
\date{\today}% It is always \today, today,

\begin{abstract}
Doppler and sideband cooling are long standing techniques that have been used together to prepare trapped atomic ions in their ground state of motion. In this paper we study how these techniques can be extended to  cool both radial modes of motion of a single ion in a Penning trap. We numerically explore the prerequisite experimental parameters for efficient Doppler cooling in the presence of an additional oscillating electric field to resonantly couple the radial modes. The simulations are supported by experimental data for a single $^{40}$Ca$^+$ ion Doppler cooled to $\sim$100 phonons in both modes at a magnetron frequency of 52 kHz and a modified cyclotron frequency of 677 kHz. For these frequencies, we then show that mean phonon numbers of $0.35(5)$ for the modified cyclotron and $1.7(2)$ for the magnetron motions are achieved after 68 ms of sideband cooling.
% \begin{description}
% Secondary publications and information retrieval purposes.
% \item[PACS numbers]
% May be entered using the \verb+\pacs{#1}+ command.
% You may use the \texttt{description} environment to structure your abstract;
% use the optional argument of the \verb+\item+ command to give the category of each item. 
% \end{description}
\end{abstract}

\pacs{Valid PACS appear here}% PACS, the Physics and Astronomy
\maketitle

%\tableofcontents
% This sample document demonstrates proper use of REV\TeX~4.1 (and
% \LaTeXe) in mansucripts prepared for submission to APS
% journals. Further information can be found in the REV\TeX~4.1
% documentation included in the distribution or available at
% \url{http://authors.aps.org/revtex4/}.

% When commands are referred to in this example file, they are always
% shown with their required arguments, using normal \TeX{} format. In
% this format, \verb+#1+, \verb+#2+, etc. stand for required
% author-supplied arguments to commands. For example, in
% \verb+\section{#1}+ the \verb+#1+ stands for the title text of the
% author's section heading, and in \verb+\title{#1}+ the \verb+#1+
% stands for the title text of the paper.

% Line breaks in section headings at all levels can be introduced using
% \textbackslash\textbackslash. A blank input line tells \TeX\ that the
% paragraph has ended. Note that top-level section headings are
% automatically uppercased. If a specific letter or word should appear in
% lowercase instead, you must escape it using \verb+\lowercase{#1}+ as
% in the word ``via'' above.
\section{\label{sec:Intro}Introduction}

Trapped ion systems provide an excellent platform for studying quantum mechanical phenomena and performing high precision measurements, due to strong isolation from the environment, high degree of control and very good spatial confinement. The latter feature is predicated on the success of various cooling techniques to reduce the motional amplitude of the ions in the trap, allowing for long interaction times and reduced Doppler shifts. In particular, for certain ion species, laser cooling techniques including resolved sideband, Raman and electromagnetically-induced transparency (EIT) methods have been able to cool single ions \cite{Diedrich1989, Monroe1995, Deslauriers2004, Che2017} or Coulomb crystals of a few to many ions to the quantum mechanical ground state of motion \cite{King1998,Lechner2016,Lin2013,Goodwin2016,Jordan2019}. In RF traps, the basic technique of a laser red-detuned from a dipole-allowed atomic resonance and passing through the trap center is sufficient for Doppler cooling of all three motional degrees of freedom of a single ion, enabling a subsequent period of sub-Doppler cooling to drive a desired mode to its  ground state. This has allowed RF trap systems to establish themselves in the field of quantum information processing \cite{Monz2015,Bermudez2017,Linke2015}, quantum simulation \cite{Kim2010,Barreiro2011,Lanyon2011,Korenblit2012,Martinez2016}, quantum optics \cite{Araneda2018} and ion clocks that employ quantum logic spectroscopy \cite{Schmidt2005}. 
\par In contrast, an ion in a Penning trap would be deconfined in the radial plane using this simple approach to Doppler cooling; instead, an offset beam geometry that uses a radial intensity gradient is required \cite{Itano1982}, resulting in an experimentally achievable parameter space that often leaves at least one of the radial motional modes with very high phonon numbers ($\bar{n}>1000$, see section \ref{sec:coolingTheory}) that are not amenable to ground state cooling techniques. For this reason, applications of Penning traps were historically mostly limited to instances where laser cooling was not required, such as mass spectrometry \cite{Blaum2010} or high precision measurements of \textit{g}-factors of electrons \cite{Hanneke2008}, protons \cite{DiSciacca2012} and anti-protons \cite{Nagahama2017}. However, to increase the precision of these measurements even further, recent experiments are being pursued where laser cooled ions are used to sympathetically cool molecular ions and highly charged atomic ions \cite{Schmidt2018}, protons and anti-protons \cite{Bohman2018} and even to perform quantum-logic spectroscopy where the electronic state of (anti-)protons can be coherently mapped and read out using an atomic ion \cite{Meiners2017}. All these approaches stand to benefit from an improved understanding of Doppler and sub-Doppler laser cooling techniques for the radial motion that will be discussed in this paper. Improved cooling also opens up potential avenues for high fidelity quantum information experiments such as Ising model quantum simulations on a 2D ion Coulomb crystal with hundreds of ions \cite{Britton2012,Bohnet2016} and implementations of error correction protocols \cite{Goodwin2015}. These experiments have motivated recent numerical study of Doppler cooling techniques for large crystals  \cite{Torrisi2016,Tang2019}. This paper complements these results by examining the case of a single-ion.
\par
This paper is split into two major sections. Section \ref{sec:DopCool} aims to find experimental parameters for which both radial modes of motion can be Doppler cooled efficiently such that subsequent resolved sideband cooling can bring the ion to the radial ground state of motion. First, a brief theoretical summary of the motional mode description and analytical Doppler cooling models highlights the underlying difficulty of laser cooling in a Penning trap. Subsequently, the axialization technique for resonant mode coupling is introduced as a means to improve the cooling limits. Finally, numerical simulations are used to find cooling limits for a variety of experimental parameters both with and without axialization. 
\par  The experimental results are then presented in section  \ref{sec:expResults}. After introducing the experimental setup, the Doppler cooling results and the effects of axialization are analyzed, showing that both modes can be cooled to $\sim$ 100 phonons. Lastly, sideband spectroscopy and Rabi oscillation data are used to demonstrate ground state cooling of both modes of motion. Section \ref{sec:conclusion} concludes the paper with a brief summary and discussion of the results.

\section{\label{sec:DopCool}Doppler cooling the radial motion in a Penning trap}
\subsection{\label{sec:level2}Motion in a Penning trap}
The classical motion of a single ion has been described in earlier works \cite{Dehmelt1968b,Brown1986}. In this section, we will summarize the key results. 

In an ideal Penning trap, a magnetic field $\textbf{B}=B\hat{\textbf{z}}$ is combined with a quadrupole electrostatic potential of the form 
\begin{equation}
V(x,y,z)= \frac{V_0}{D_0^2}(2z^2-x^2-y^2),
\end{equation}
where $D_0$ characterizes the geometric dimension of the trap and $V_0$ is the applied voltage. A particle with mass $M$ and and charge $q$ will thus experience a force
\begin{equation}
M\ddot{\textbf{r}} = q\dot{\textbf{r}}\times \textbf{B}-q\nabla V(\textbf{r}).
\label{eq:potential}
\end{equation}
The solutions of the equations of motions in terms of the Cartesian coordinates $(x,y,z)$ are
\begin{subequations}
    \begin{align}
    x(t) &=  r_-\cos(\omega_-t+\phi_-)+r_+\cos(\omega_+t+\phi_+),\\
    y(t) &= - r_-\sin(\omega_-t+\phi_-)-r_+\sin(\omega_+t+\phi_+),\\
    z(t) &= r_z\cos(\omega_zt+\phi_z),
    \end{align}
\label{eq:eqMotion}
\end{subequations}
resulting in three motional modes with amplitudes $r_{\pm},r_{z}$ and phases $\phi_{\pm},\phi_{z}$, which depend on the initial conditions of position and velocity.
The presence of the $\textbf{B}$ field couples the motion in the radial $xy$ plane, resulting in mode frequencies
\begin{equation}
\omega_{\pm}=\frac{1}{2}\left(\omega_c\pm\sqrt{\omega_c^2-2\omega_z^2}\right)=\frac{\omega_c}{2}\pm\omega_1,
\label{eq:radFreqs}
\end{equation}
where $\omega_c=qB/M$ is the true cyclotron frequency and $\omega_1=\sqrt{\omega_c^2-2\omega_z^2}/2$. In the radial plane, the particle thus undergoes a superposition of circular motions at the modified cyclotron $\omega_+$ and magnetron $\omega_-$ frequencies, whereas in the axial ($z$) direction the motion is simple harmonic with frequency $\omega_z=\sqrt{4qV_0/(MD_0^2)}$. The stability limit of the trap is set by $\sqrt{2}\omega_z\leq\omega_c$ such that the frequencies in equation \ref{eq:radFreqs} remain real. This implies that for a given magnetic field, the applied voltage is limited to: $V_0\leq qD_0^2B^2/(8M)$. 
\par 
Next, we find the cycle averaged kinetic energy and potential energy using the results from equation \ref{eq:eqMotion}
\begin{align}
\langle E_K \rangle = & \langle \frac{1}{2}M\dot{\textbf{r}}^2 
\rangle=  \frac{1}{2}M\left(\frac{1}{2}r_z^2\omega_z^2+r_+^2\omega_+^2+r_-^2\omega_-^2\right), \label{eq:kinEn}\\
\langle E_V \rangle = & \langle qV\rangle= \frac{1}{4}M\omega_z^2\left(r_z^2-r_+^2-r_-^2\right).
\label{eq:potEn}
\end{align}
The total energy of motion is thus the sum of equations \ref{eq:kinEn} and  \ref{eq:potEn}
\begin{equation}
\langle E_T\rangle=\frac{1}{2}M\left(r_z^2\omega^2_z+2r_+^2\omega_+\omega_1-2r^2_-\omega_-\omega_1\right).
\label{eq:totEnergy}
\end{equation}
From equation \ref{eq:totEnergy} we see that the term with the magnetron frequency is negative. This means that any reduction in the amplitude, and hence the kinetic energy, of the magnetron mode $r_-$  actually increases the total energy. Furthermore, the motion is now also unstable against perturbations, such as background gas collisions, that increase $r_-$. Thus to effectively cool the particle, defined as a reduction of its kinetic energy irrespective of the total energy, the magnetron mode of motion will require a cooling mechanism that doesn't simply dissipate energy. The implications of this will be further discussed in the next section. 
\par 
The quantized motion of the Penning trap can be formulated by imposing the canonical commutation relations $[\hat{q}_i,\hat{p}_j]=i\hbar\delta_{ij}$ on the classical canonically conjugate variables $q_i$ and $p_j$. Using an appropriate coordinate transformation \cite{Brown1986,Kretzschmar1991,Crimin2018}, the Hamiltonian takes the form of three uncoupled harmonic oscillators
% \begin{equation}
%     \hat{H}=\hbar\omega_z\left(\hat{a}^{\dagger}_z\hat{a}_z+\frac{1}{2}\right)+\left(\hat{a}^{\dagger}_+\hat{a}_++\frac{1}{2}\right)-\left(\hat{a}^{\dagger}_-\hat{a}_-+\frac{1}{2}+\right)
% \end{equation}
 \begin{equation}
    \hat{H}=\hbar\omega_z\left(\hat{n}_z+\frac{1}{2}\right)+\hbar\omega_+\left(\hat{n}_++\frac{1}{2}\right)-\hbar\omega_-\left(\hat{n}_-+\frac{1}{2}\right),
    \label{eq:penningHamil}
 \end{equation}
 in terms of the number operators ${\hat{n}_z,\hat{n}_+,\hat{n}_-}$ for the axial, modified cyclotron and magnetron modes respectively. By defining $r_0=\sqrt{\frac{\hbar}{M\omega_1}}$ and $z_0=\sqrt{\frac{\hbar}{M\omega_z}}$ as the length scale of the ground state wave-packet of the radial and axial motion respectively and equating eq. \ref{eq:totEnergy} to eq. \ref{eq:penningHamil} the classical amplitude of motion can be related to the occupation number of each mode:
 \begin{subequations}
    \begin{align}
     r_z^2 &= 2z_0^2\left(n_z+\frac{1}{2}\right),\\
     r_+^2 &= r_0^2\left(n_++\frac{1}{2}\right),\\
     r_-^2 &= r_0^2\left(n_-+\frac{1}{2}\right).
    \end{align}
\end{subequations}
\subsection{\label{sec:coolingTheory}Cooling Theory}
\subsubsection{Offset beam cooling}
The simplest laser cooling technique typically uses a laser beam that is red-detuned from a strong dipole-allowed atomic transition to create a velocity selective damping force that removes energy from the particle. In the case of a harmonically trapped two-level particle in one dimension, photons are preferentially scattered when the ion is moving towards a laser beam parallel to the trap axis, as they come into resonance when the Doppler frequency shift approaches the laser detuning. After many such scattering events, the Doppler limited temperature, assuming the photon emission to be isotropic, is given by \cite{Itano1982}: $ T=\hbar\Gamma/(3k_B)$,
%\begin{equation}
%    T=\frac{\hbar\Gamma}{3k_B},
%\end{equation}
where $\Gamma\ $is the transition linewidth and $k_B$ is the Boltzmann constant. In a multi-level system, such as the $^{40}$Ca$^+$ ion, the particle can decay to a long-lived metastable state from which it is returned back into the cooling cycle with the aid of an additional repump laser. The effect of this interaction, away from potential dark resonances, is taken to be small (e.g. $\sim$ 6$\%$ branching ratio in $^{40}$Ca$^+$) and for the purposes of this discussion ignored. For a harmonically bound particle in three dimensions in a Paul trap, a single laser aimed at the trap center with a projection onto all principal axes of motion is sufficient to cool all modes of motion. In a Penning trap, the same geometry would cool the axial $\omega_z$ mode and the radial $\omega_+$ modified cyclotron mode, whilst heating the $\omega_-$ magnetron mode. This is a direct consequence of the negative total energy seen in equation $\ref{eq:totEnergy}$, which causes a sign reversal in the rate of change of kinetic energy between $\omega_+$ and $\omega_-$. Thus to cool the magnetron motion, a blue-detuned laser beam would be required, implying there is no frequency selective force that can cool both modes simultaneously \cite{Horvath1999}.
\par
\begin{figure}[h]
\centering
\includegraphics[width=0.9\linewidth]{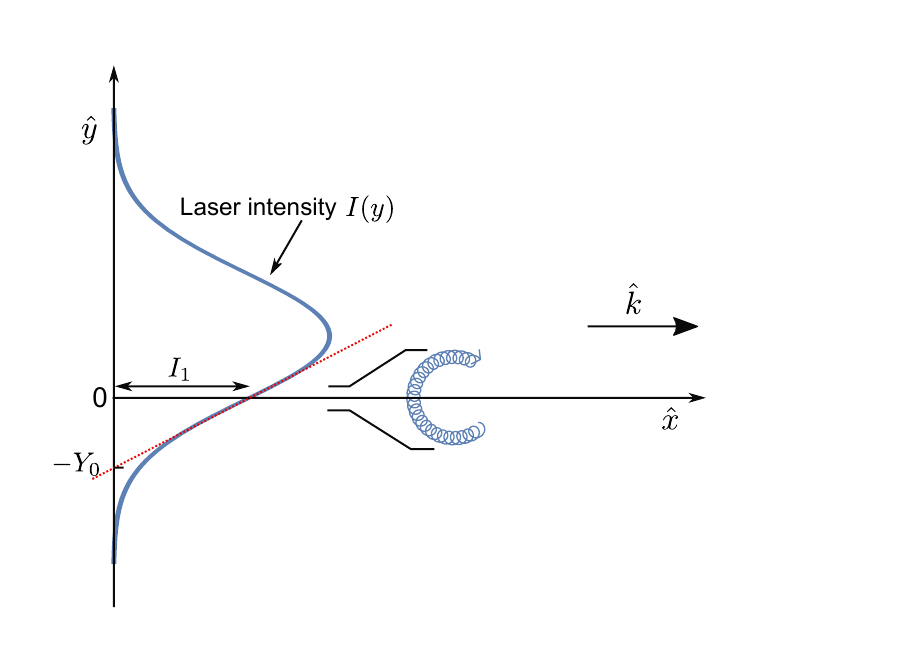} 
\caption{Schematic diagram of the offset Doppler cooling beam relative to the ion trajectory in the trap. The clockwise trajectory is exaggerated for clarity. }
\label{fig:beamCoolingSchematic}
\end{figure}

\begin{figure}[h]
\centering
\includegraphics[width=1\linewidth]{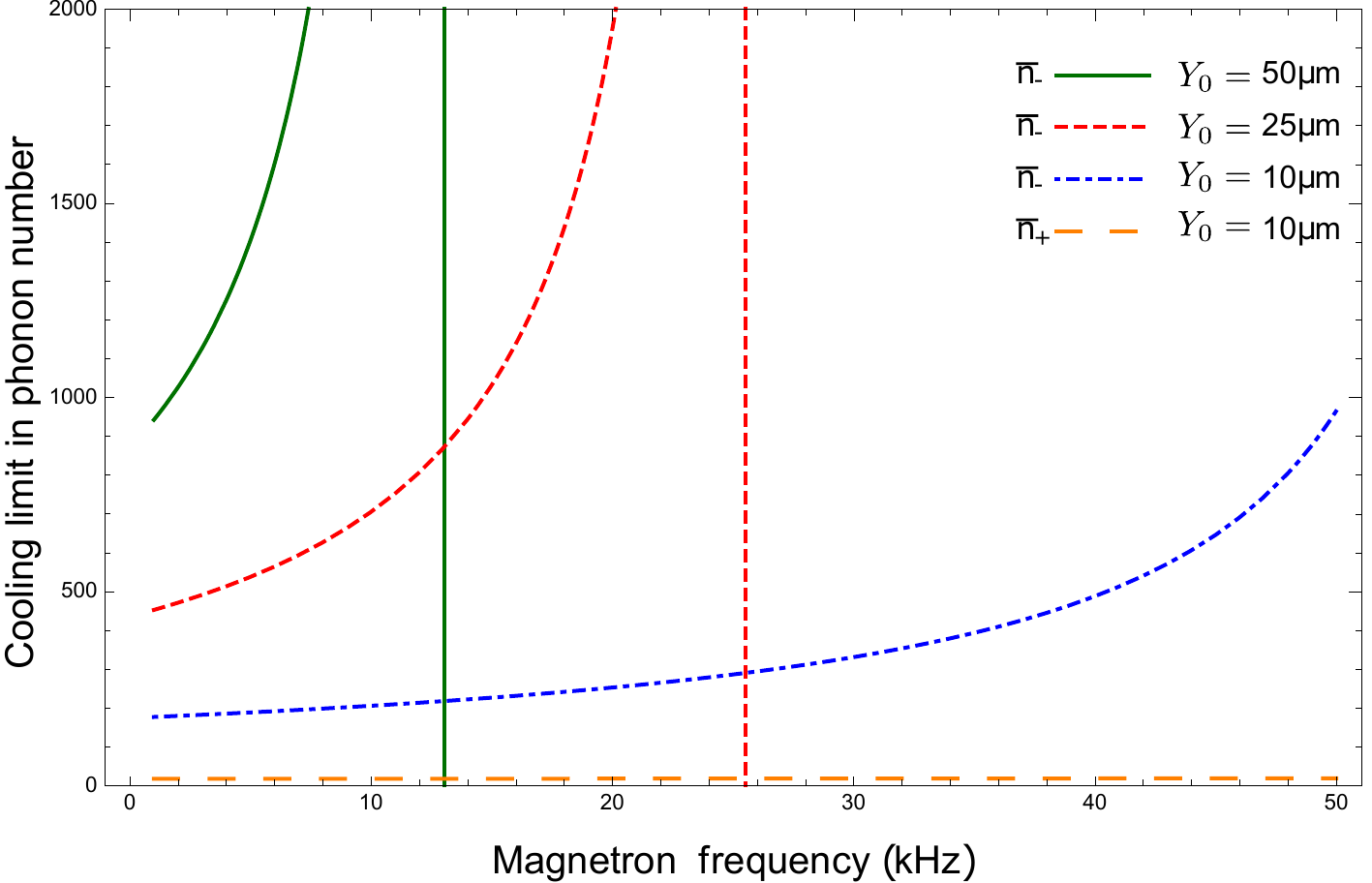} 
\caption{Doppler cooling limit in terms of mean phonon number as a function of magnetron frequency for different beam offset parameters. The vertical lines for the red and green curves intersecting zero show the limit of cooling
stability. Note that the modified cyclotron cooling limit response is almost flat even for a small offset parameter, in contrast with the rapid growth of the magnetron cooling limit for larger offset parameters. }
\label{fig:analyticCoolLim}
\end{figure}

 
To achieve simultaneous cooling of both radial modes, a spatial degree of freedom must be introduced, whereby an inhomogeneous laser intensity profile breaks the symmetry around the center of the trap, leading to enhanced scattering when the ion is in the half of the magnetron orbit that moves away from the laser beam. A typical laser beam with a Gaussian intensity profile:
\begin{equation}
    I=\frac{2P_0}{\pi w_0^2}e^{-2(y-y_0)^2/w_0^2},
\end{equation}
with power $P_0$, offsetting beam waist $w_0$ offset from the trap center by $y_0$ creates a scattering rate gradient across the ion trajectory as shown in figure \ref{fig:beamCoolingSchematic} and hence creates a position dependent damping force for the magnetron motion. To facilitate an analytic derivation of the cooling limits, the beam intensity can be approximated as a linear function of the ion position $y$ giving $I(y)=I_1(1+y/Y_0)$ with $I_1$ being the intensity at the trap centre, provided that the ion orbit is small compared to the gradient parameter $Y_0$, i.e.  $\lvert y\rvert\ll Y_0$. The largest gradient is obtained for an offset of $y_0=w_0/2$ hence, for a given beam waist, the optimal gradient parameter is $Y_0=w_0/2$. The initial theoretical analysis of Itano and Wineland \cite{Itano1982} determined that to achieve simultaneous cooling of both modes an inequality must be obeyed:
\begin{equation}
    \omega_-<\frac{(\Gamma/2)^2+\delta^2}{2kY_0\delta}<\omega_+,
\end{equation}
where $\Gamma$ is the transition linewidth, $\delta$ is the laser detuning (defined as positive below resonance) and $k$ is the wavenumber of the laser beam. This limit can be viewed as the parameter space for which simultaneous cooling of both modes can be obtained, but does not constitute a hard cut-off since the approximation breaks down as the ion orbit grows. More recent analysis used numerical simulations to show that there are large amplitude steady states when accounting for the full laser profile \cite{Horvath1999}. It is also possible to derive analytical expressions for the cooling rates as a function of laser beam offset and detuning, including transition saturation effects \cite{Thompson2000}.
Nonetheless, the resulting approximate mean phonon number expressions at the Doppler limit remain useful and are given for both motions by \cite{Itano1982}:
\begin{align}
    \bar{n}_+\approx\frac{5Y_0 k((\Gamma/2)^2+\delta^2)}{6[2\delta\omega_+Y_0k-((\Gamma/2)^2+\delta^2)]}  \label{eq:DopplerLimC}\\
    \bar{n}_-\approx\frac{5Y_0 k((\Gamma/2)^2+\delta^2)}{6[((\Gamma/2)^2+\delta^2)-2\delta\omega_-Y_0k]}
    \label{eq:DopplerLimM}
\end{align}
assuming no axial cooling beam and isotropic emission. An important thing to note from equations \ref{eq:DopplerLimC} and \ref{eq:DopplerLimM} is that the sign of the position and detuning dependent terms in the denominator is reversed showing the conflicting requirements on $\delta$ and $Y_0$ to minimise the respective mean phonon numbers. Furthermore, since $\omega_-$ can be up to two orders of magnitude smaller than $\omega_+$, the corresponding Doppler limit can be much higher. To illustrate this, figure \ref{fig:analyticCoolLim} shows the Doppler limits at $\delta=\Gamma/2$ for a $^{40}$Ca$^+$ ion trapped in the Imperial College Penning trap \cite{Mavadia2013} operated at a $1.89$ T magnetic field. It is clear that to achieve sufficiently low mean phonon numbers of the magnetron motion at frequencies that are amenable to resolved sideband cooling, the $Y_0$ parameter needs to be as small as possible. For example, a mean phonon number of $\bar{n}_-\approx 200$ at $\nu_z=350$ kHz requires a beam waist of 20 $\mu$m. This is often in conflict with other experimental requirements. A tight waist is not suitable for initial cooling from large magnetron orbits and hence would greatly complicate the trap loading process. Even if an ion is successfully cooled to the trap center, any strong perturbation such as an elastic collision with a background gas molecule would leave the ion in a large orbit from which it can not be efficiently re-cooled. Furthermore, a tight waist becomes very sensitive to beam pointing instabilities and makes it difficult to achieve a reproducible Doppler temperature without additional stabilization techniques. Lastly, a tight waist is also not suitable for cooling larger Coulomb crystals, which can have radii from tens to hundreds of $\mu$m. For these reasons, offset beam cooling alone is not sufficient to reach a suitably low mean phonon occupation to enable sideband cooling to the ground state of the magnetron motion.
%Since the goal of all cooling techniques is to ultimately reduce Doppler s, by rec to cooling as the reduction of kinetic energy without considering the changes to the total energy. We can now begin to see why cooling is more complicated in a Penning trap compared to 

\subsubsection{Axialization}
In the previous section we showed that the two radial modes of motion can have vastly different Doppler cooling limits. One method to bridge this difference was originally proposed by Dehmelt and Wineland \cite{Wineland1975338}: a resonant coupling between the axial and magnetron motion would lead to continuous exchange of energy between them and if the cooling of the axial motion was faster than the heating of the magnetron motion, the two modes could be cooled simultaneously. This technique was initially demonstrated in a high precision experiment which measured the anomalous magnetic moment of the electron in a Penning trap \cite{VanDyck1977}, and was later used also used on ions \cite{Cornell1990}. Similarly, a coupling between the modified cyclotron and magnetron motions can be achieved using an azimuthal quadrupolar field as produced with a ring electrode split into four segments \cite{Bollen1990}. This effect is achieved by applying an oscillating quadrupolar potential:
\begin{equation}
    \phi_{ax}=\frac{V_a}{2R_0^2}(x^2-y^2)\sin(\omega_ct),
\end{equation}
with voltage $V_a$ at the true cyclotron frequency $\omega_c=\omega_++\omega_-$ to the segments of a ring electrode of effective radius $R_0$ with each neighbouring segment phase shifted by $\pi$ radians. Combining this potential with the radial equations of motion, equation \ref{eq:potential} yields:
\begin{equation}
\label{eq:axMotion}
    \ddot{u}+i\dot{u}\omega_c-\left(\frac{\omega_z^2}{2}-i\frac{qV_{ax}}{MR_0^2}\sin(\omega_ct)\right)=0,
\end{equation}
where $u=x+iy$. This equation can be solved either by considering the instantaneous power absorption \cite{Schweikhard1993} or Green's functions \cite{Cornell1990,Brown1986}. The solutions yield a simple sinusoidal coupling at a rate:
\begin{equation}
    \Omega_a=\frac{qV_a}{4MR_0^2\omega_1}.
\end{equation}
  For a 1 volt drive on a $^{40}$Ca$^+$ ion with $R_0=0.01$m, $\omega_1=2\pi\times 300$ kHz, the coupling rate is $\Omega_a=2\pi\times 500$ Hz, which is much slower than all the motional frequencies of the ion, allowing for a few complete energy exchange cycles during a typical Doppler cooling window (5-20 ms). In the presence of efficient damping of one of the modes (by laser cooling or otherwise), the axialization will drive the mean phonon numbers of the coupled modes towards equilibrium, $\bar{n}_+=\bar{n}_-$ \cite{Wineland1979,Brown1986}, however establishing what is the equilibrium value is not trivial.
% \begin{figure}[h]
% \centering
% \includegraphics[width=1\linewidth]{AxiTrajectory.pdf} 
% \caption[Average phonon numbers vs. axialization amplitude]{Average phonon numbers for the magnetron (circles) and modified cyclotron (diamonds) motions between 10--20 ms as a function of axialization voltage for different beam waists. Parameters: $\nu_z=265$ kHz, $\nu_-=52$ kHz, $\nu_+=677$ kHz, $y_0= 0$ $\mu$m, $P_0=8$ $\mu$W,  $\delta=\Gamma/2$. }
% \label{fig:AxiTraj}
% \end{figure}
 \par Experimentally, the axialization technique was demonstrated to successfully cool the magnetron motion of $^{24}$Mg$^+$ ions \cite{Powell2002,Powell2003} and single $^{40}$Ca$^+$ ions \cite{Goodwin2016}, clouds \cite{Phillips2008} and crystals \cite{Crick2008,Stutter2018} compared to an offset beam-only method, however no direct thermometry was performed. The thesis of Stutter \cite{Stutter2015} includes a brief study of a radial spectrum in the presence of axialization, but observed high order sideband excitations (8-10th order) which were attributed to signatures of large driven motion orbits, hence no temperature was derived.  Theoretical studies have so far focused on the cooling rates \cite{Hendricks2008a} without consideration of the final cooling limits. In the following section we  numerically investigate cooling limits in the presence of axialization and compare them with experimental results.

% The \texttt{widetext} environment will make the text the width of the
% full page, as on page~\pageref{eq:wideeq}. (Note the use the
% \verb+\pageref{#1}+ command to refer to the page number.) 
% \paragraph{Note (Fourth-level head is run in)}
% The width-changing commands only take effect in two-column formatting. 
% There is no effect if text is in a single column.

\subsection{\label{sec:DopplerSim}Doppler cooling simulation and results}
\subsubsection{Method}
To investigate the cooling limits we numerically integrate the radial equations of motion (eq. \ref{eq:axMotion}) in the presence of stochastic momentum kicks from the cooling laser. To model the laser-ion interaction, we  use a semi-classical model where we assume that a monochromatic laser photon with wavevector $\mathbf{k}$ will cause an instantaneous velocity change $\Delta \mathbf{v}=\hbar(\mathbf{k}'-\mathbf{k})/M$ of the ion after re-emission with wavevector $\mathbf{k}'$. This approximation is valid since the lifetime of the excited state of the cooling transition is much shorter than any of the oscillation periods of the ion motion. The laser linewidth is much smaller than the transition linewidth $\Gamma$, satisfying the monochromatic assumption and the beam is normal to the magnetic field axis with $\mathbf{k}=k\Hat{\mathbf{x}}$. 
%, frequency $\omega_L$ and detuning $\delta$
%=\mathbf{v}-\mathbf{v}'.
The scattering rate,  including the Doppler shift, is now given by:
\begin{equation}
\gamma_g(t)=\frac{I(y)\sigma_{0}}{\hbar\omega_L}\frac{(\Gamma/2)^2}{(\Gamma/2)^2+\frac{I(y)\sigma_{0}\Gamma}{2\hbar\omega_L}+(\delta+\dot{x}(t)k)^2},
\label{eq:scatterRateRad}
\end{equation}
where $\sigma_{0}$ is the scattering cross-section. In our experiment we address both $\sigma^{\pm}$ transitions with linear polarization and hence obtain $\sigma_0=\lambda^2/(2\pi)$.
Finally, we assume that the photon statistics are Poissonian and thus we can define a probability of scattering integer $n$ photons in a time interval $dt$ as $P_n(\gamma_g(t)dt)$. 
\par
The integration algorithm then proceeds as:
\begin{enumerate}
\item Integrate the equations of motion (eq. \ref{eq:potential}) using an 8\textsuperscript{th} order Runge-Kutta method by a time step $dt$. 
\item Sample the Poissonian distribution $P_n(\gamma_g(t)dt)$ to find how many photons $n$ arrived in the time step. 
\item Generate $n$ uniformly distributed random vectors on a sphere of radius $k$ and sum them to find the total re-emission wave vector $\textbf{k}'$.
\item Update the velocity coordinates by the corresponding velocity kicks:
\begin{align}
\dot{x}(t)&\rightarrow\dot{x}(t)+\frac{\hbar}{M}(kn+k'_x),\\
\dot{y}(t)&\rightarrow\dot{y}(t)+\frac{\hbar}{M}k'_y.
\end{align}
\end{enumerate}
The instantaneous square amplitudes of motion can be found by substituting the integrated coordinates into the cycle averaged analytical square amplitudes:
\begin{align}
\label{eq:cycAmp}
\langle r_+^2\rangle=&\frac{1}{4\omega_1^2}\left[(\omega_-x(t)+\dot{y}(t))^2+(\omega_-y(t)-\dot{x}(t))^2\right],\\
\label{eq:magAmp}
\langle r_-^2\rangle=&\frac{1}{4\omega_1^2}\left[(\omega_+x(t)+\dot{y}(t))^2+(\omega_+y(t)-\dot{x}(t))^2\right].
\end{align}
The mean phonon numbers can then be expressed in terms of the amplitudes as: 
\begin{equation}
\bar{n}_{\pm}=\frac{\langle r_{\pm}^2\rangle M\omega_1}{\hbar},
\label{eq:meanPhonon}
\end{equation}
where we have neglected the zero-point energy. An important point to note is that the amplitudes are defined with respect to the center of the trapping potential, however, since the laser exerts a constant force in the $\hat{x}$ direction, it displaces the ion trajectory from the center, resulting in oscillatory values of $\langle r_+^2\rangle$ and $\langle r_-^2\rangle$ at the magnetron frequency $\omega_-$. If the cooling laser is turned off non-adiabatically, then the displacement is imprinted on the measured amplitude in each experimental realization depending on the phase of the oscillation. While this displacement increases or decreases the total with equal probability, the mean phonon number increases on average since it depends on the square of the displacement (eq. \ref{eq:meanPhonon}). This process results in the creation of non-thermal, coherent states of motion.  Since the magnitude of this heating varies with trap frequency and scattering rate, the position offset due to the constant force term, $2\hbar k \gamma_g(t_0)/(M\omega_z^2)$, is subtracted from the $x$-coordinates of the solution when calculating $\langle r_+^2\rangle$ and $\langle r_-^2\rangle$ to ensure consistency between simulated data sets. Experimentally, this heating process can be suppressed by turning off the cooling laser beam adiabatically with respect to the magnetron frequency.  
\par 
The final issue to consider is how to extract equilibrium mean phonon values from the resulting trajectories. Given a particular sets of parameters, a trajectory solution might show oscillatory behaviour or very slow convergence to equilibrium. In a real experiment, cooling is typically performed for a fixed time period after which spectroscopy would be performed. To restrict the simulation to realistic parameters that show fast dynamical convergence to the cooling limit, can overcome heating and are robust at re-cooling ions that have collided with background particles, appropriate integration limits are chosen. The equations of motion are integrated for 20 ms and the amplitudes are averaged over a 10-20 ms interval assuming that the ergodicity of a single trajectory gives us information about the average ensemble of amplitudes. The same initial conditions are used for all simulations: $x_i=-4\times10^{-6}$ m, $v_x=1$ m\,s$^{-1}$ and $v_y=2$ m\,s$^{-1}$. This corresponds to initial phonon numbers of $\bar{n}_-\approx 18250 \pm 150$, $\bar{n}_+\approx 200 \pm 100$ across the simulated frequency range. 
\subsubsection{Results}
\begin{figure}[h]
\begin{subfigure}[b]{0.45\textwidth}
\centering 
\includegraphics[width=1\linewidth]{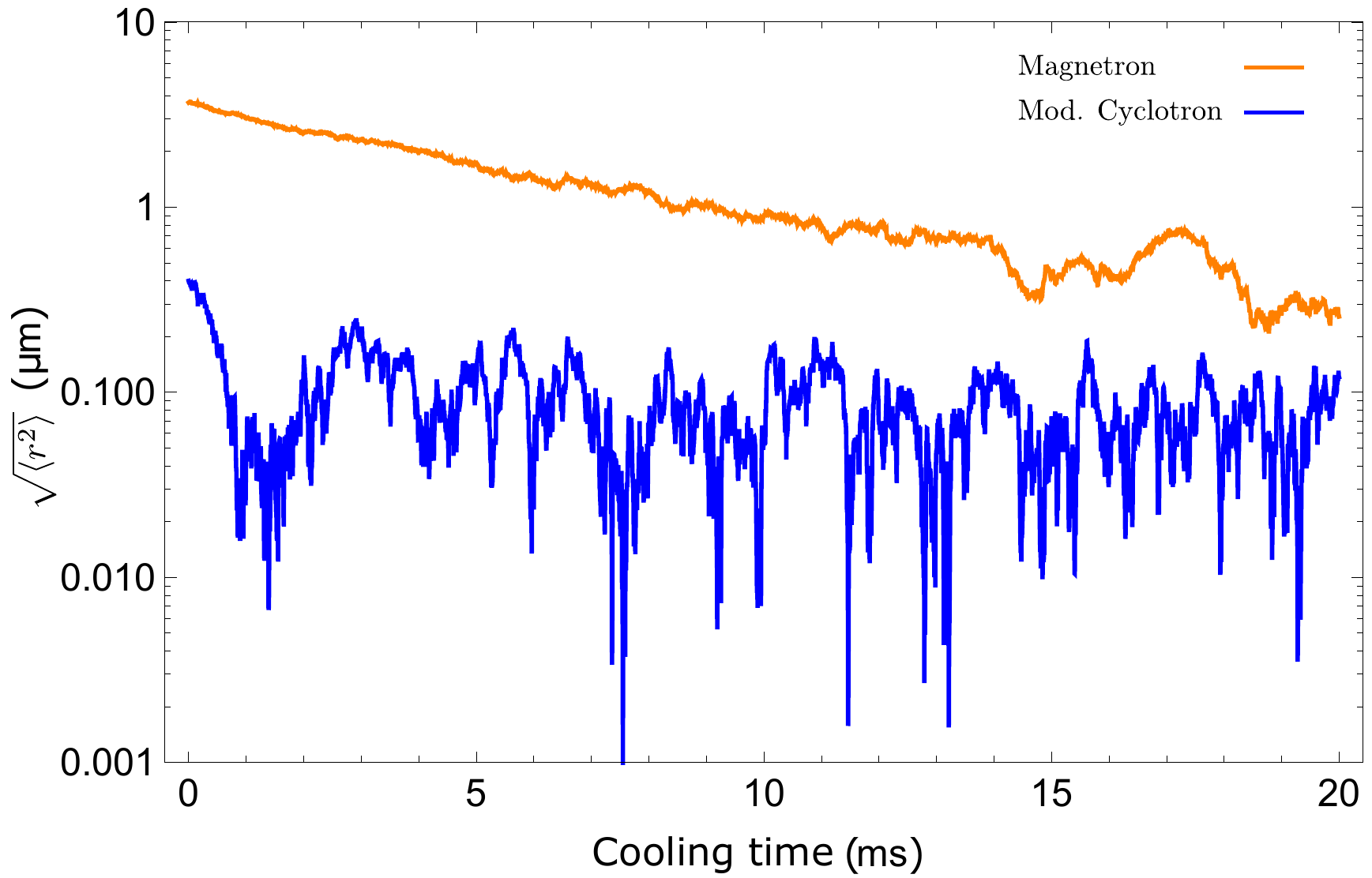}
\caption{$V_a=0$V}
\end{subfigure}
\centering
\begin{subfigure}[b]{0.45\textwidth}
\includegraphics[width=1\linewidth]{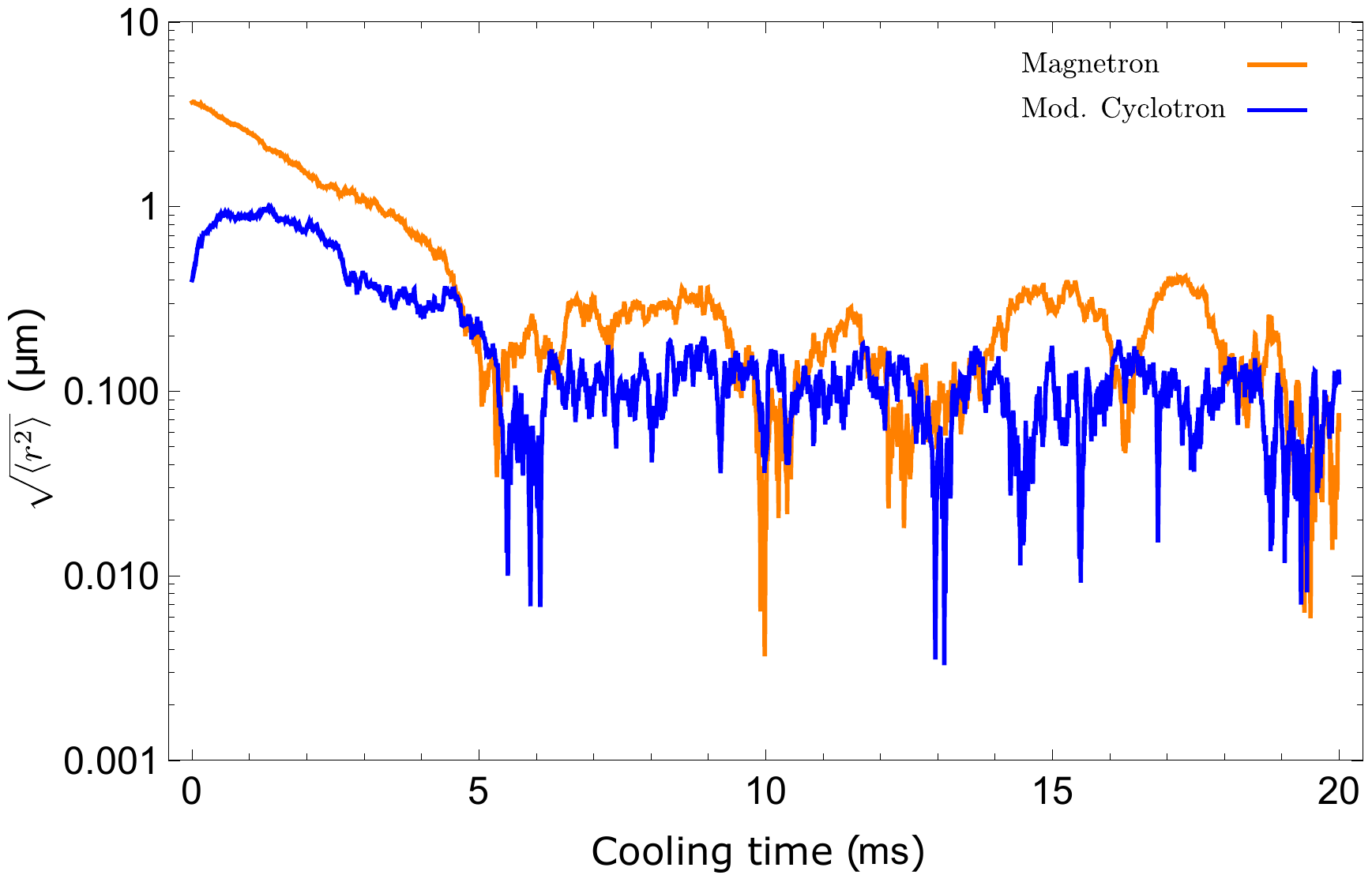}
\caption{$V_a = 0.5$V}
\end{subfigure}
\caption{Amplitudes of motion for the magnetron (orange) and modified cyclotron (blue). In figure (b) the characteristic initial increase in the cyclotron amplitude due to the presence of the axialization drive is observed. This effect is absent in figure (a) where no axialization is present.  Common parameters: $P_0= 1$ $\mu$W, $\delta=\Gamma/2$, $w_0= 40$ $\mu$m, $y_0 = 40$ $\mu$m,  $\nu_-=25$ kHz }
\label{fig:trajectoryPlots}
\end{figure}

We begin by showing the results of typical trajectory data used to extract equilibrium phonon numbers. In figure \ref{fig:trajectoryPlots}(a), the amplitudes of both motions are cooled by means of an offset Doppler beam without axialization. The modified cyclotron amplitude decreases rapidly within hundreds of $\mu$s, while the magnetron amplitude cools much slower and does not reach equilibrium in 20 ms. In figure \ref{fig:trajectoryPlots}(b), the modified cyclotron amplitude initially increases due to the presence of 0.5 V of axialization in addition to an offset Doppler beam, while cooling the magnetron motion. These axialization induced oscillations disappear once equilibrium is reached and the mean phonon number can once again be extracted by averaging over the 10-20 ms period where the amplitude is dominated by random kicks from the photon scattering. 
\par
\begin{figure}[h]
\centering
\includegraphics[width=1\linewidth]{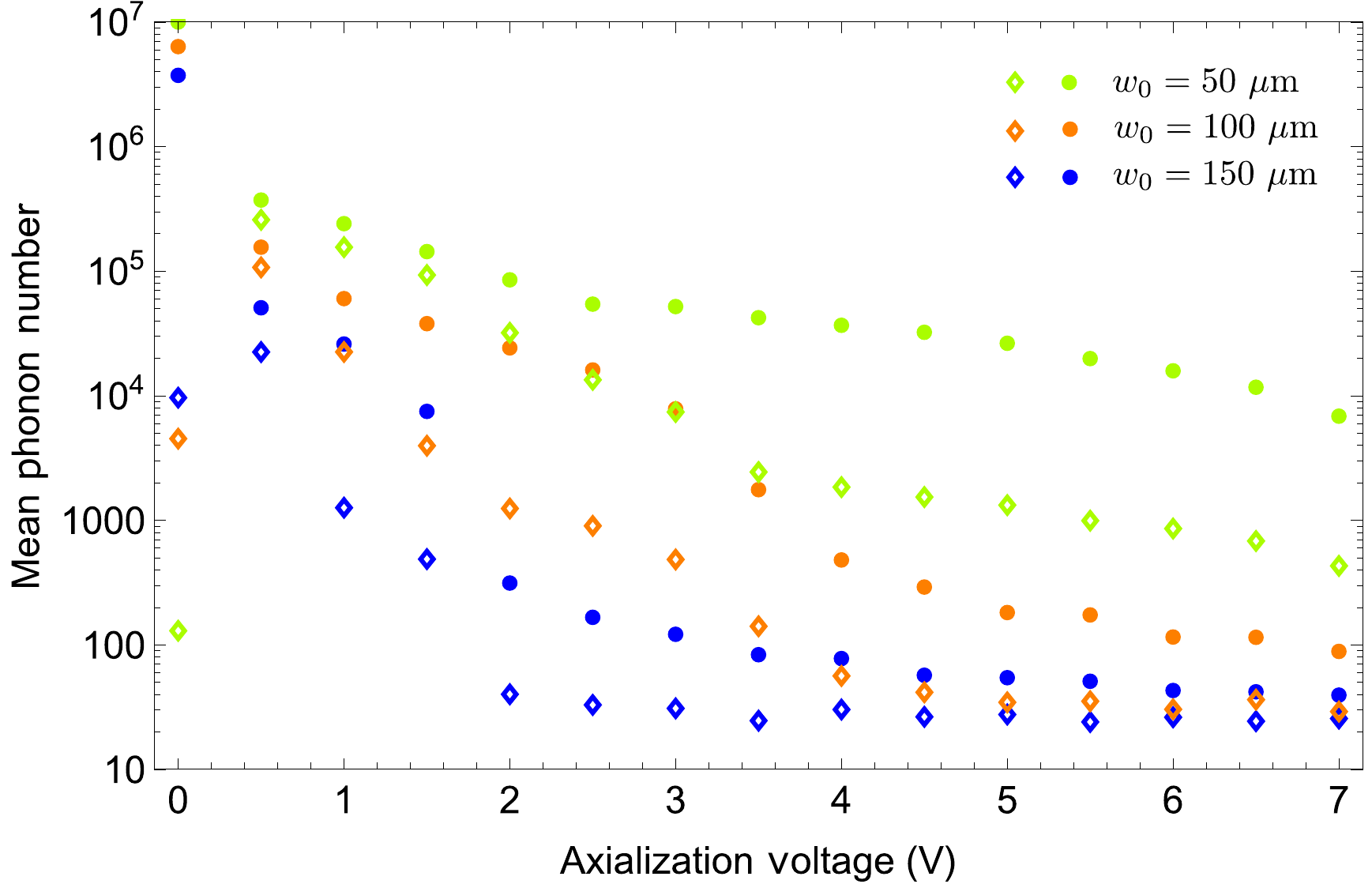} 
\caption[Average phonon numbers vs. axialization amplitude]{Average phonon numbers for the magnetron (circles) and modified cyclotron (diamonds) motions between 10--20 ms as a function of axialization voltage for different beam waists. Parameters: $\nu_z=265$ kHz, $\nu_-=52$ kHz, $\nu_+=677$ kHz, $y_0= 0$ $\mu$m, $P_0=8$ $\mu$W,  $\delta=\Gamma/2$. }
\label{fig:axiNoOffsetFinal}
\end{figure}

 
Based on the technique for extracting equilibrium phonon numbers described above, the effect of cooling using axialization without an offset beam was investigated. Figure \ref{fig:axiNoOffsetFinal} plots various cooling limits as a function of the axialization voltage for different beam waists. First, note that at $V_{ax}=0$ the magnetron motion is actively heated to very large orbits while the modified cyclotron motion remains relatively cold, consistent with the theoretical prediction that both modes cannot be cooled simultaneously. Furthermore, the larger scattering rate for smaller beam waists at the trap center leads to more efficient modified cyclotron cooling. As the axialization amplitude is increased, the magnetron mean phonon numbers start decreasing. The modified cyclotron mean phonon numbers initially go up, but with sufficient axialization the two values get closer to each other and approach a low total phonon number. It is clear that lower total mean phonon numbers can be obtained for larger beam waists, indicating that the broader beam waist contributes a larger net decrease in magnetron heating compared to the decrease in modified cyclotron cooling. Phonon numbers below 100 for both modes can be obtained by using around $V_{ax}>3$ V for a beam waist of $w_0=150 $ $\mu$m, which would be sufficient for commencing sideband cooling. The disadvantage of using voltages in this range is that driven micromotion can become significant if stray fields and potential differences between the axialization voltages on different electrodes are not carefully compensated. 

\par
\begin{figure}[h]
\centering
\includegraphics[width=1\linewidth]{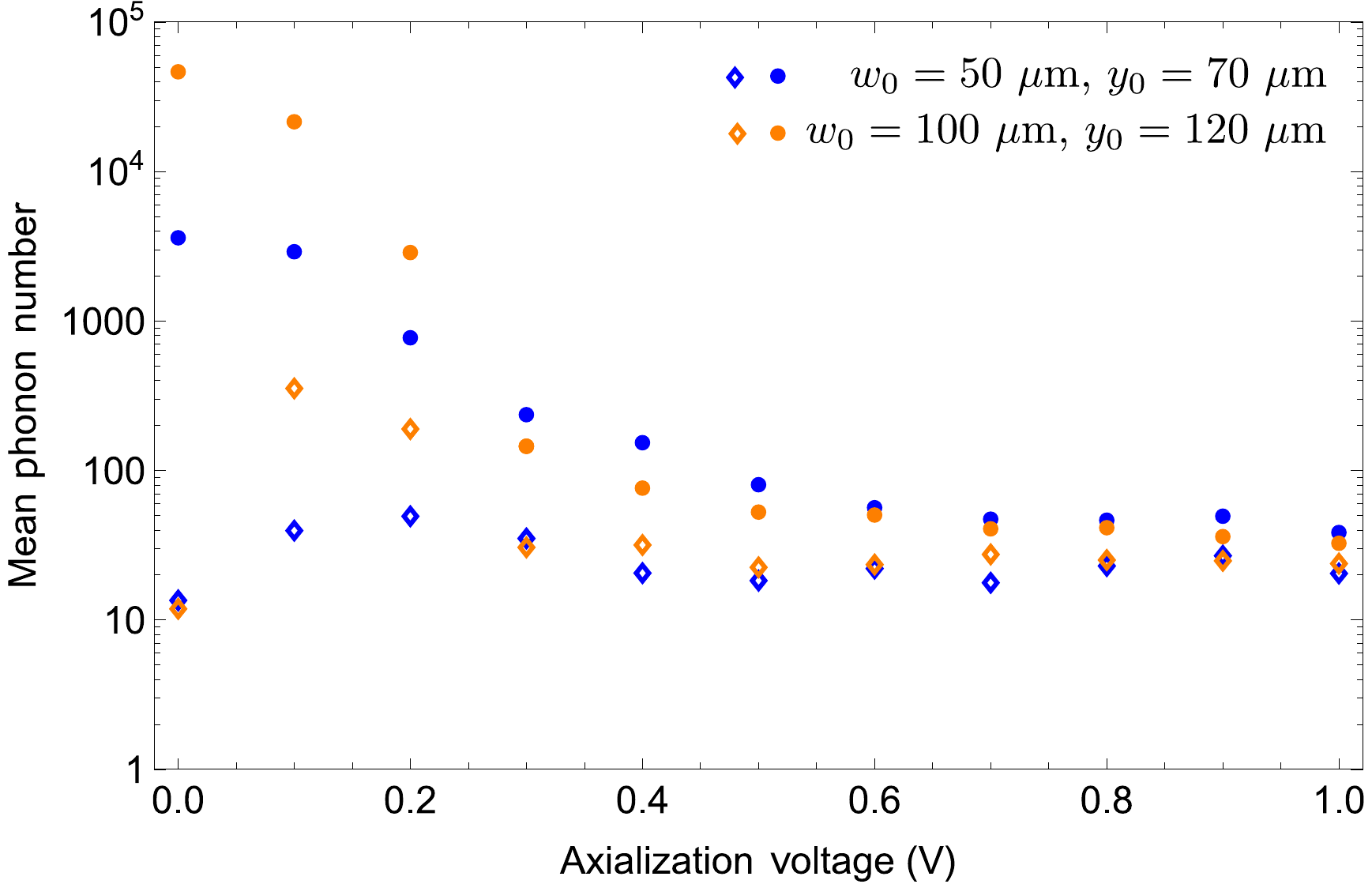} 
\caption[Average phonon numbers vs. Axialization amplitude with offset beam]{Average phonon numbers for the magnetron (circles) and modified cyclotron (diamonds) motions between 10--20 ms as a function of axialization voltage for different beam waist and offset combinations. The offset was chosen to minimize the mean phonon numbers given the beam waist and power based on additional simulations at $V_{ax}=0.5$V. Parameters: Trap potential 100 V, $\nu_z=265$ kHz, $\nu_-=52$ kHz, $\nu_+=677$ kHz, $P_0=8$ $\mu$W, $\delta=\Gamma/2$.}
\label{fig:axvsVolt}
\end{figure}

 
An alternative is to combine axialization with an offset Doppler cooling beam as shown in figure \ref{fig:axvsVolt} for two different beam waists. In this case, the axialization voltages required to achieve similar mean phonon numbers is an order of magnitude lower. At around $V_{ax}=0.7$ V the effect of additional voltage starts to saturate and the final mean phonon numbers of both modes are almost equalized. The phonon numbers are within a factor of 3-4 larger than the modified cyclotron mean phonon number without axialization. Additionally, the benefits of a tighter beam waist are also evident at lower axialization voltages. 
\par 
Further simulations using axialization voltages of $V_{ax}=0.5$V and $V_{ax}=1$V show that the mean phonon numbers of the two motions are almost equalized at values between 10 to 200 phonons at magnetron frequencies from $\nu_-=10$ to $\nu_-=190$ kHz. This very weak dependence on the applied DC endcap voltage, suggests that with sufficient axialization voltage, efficient cooling is possible across the whole range of stable trap parameters.
%The final result of this section in figure \ref{fig:axiTrapVoltagePlot} shows the dependence of the mean phonon number as a function of the axial trapping frequency in the presence of different axialization voltages for the beam parameters used in our experiment. For $V_{ax}=0$V the $\bar{n}_-$ rises rapidly, while $\bar{n}_+$ has only a very weak dependence. By $\nu_z=300$ kHz the ion is heated to a very large orbit where it only weakly interacts with the laser when reaching a steady state. A similar trend is observed for $V_{ax}=0.1$V, but there is now evidence of the steady state orbit of the magnetron motion being an order of magnitude smaller at the expense of slightly higher $\bar{n}_+$, showing a clear net reduction of the combined phonon numbers. For higher axialization voltages of $V_{ax}=0.5$V and $V_{ax}=1$V, the mean phonon numbers of the two motions are almost equalised and show only a very weak dependence on the axial frequency, suggesting that with sufficient axialization voltage efficient cooling is possible across the whole range of stable trap parameters.
%\begin{figure}[!h]
%\centering
%\includegraphics[width=1\linewidth]{axiTrapVoltagePlotPRA.pdf} 
%\caption[Average phonon numbers vs. axial frequency]{Average phonon numbers for the magnetron (circles) and modified cyclotron (diamonds) motions between 10--20 ms as a function of axial frequency $\nu_z$ for different axialization voltages. Parameters: $w_0=100$ $\mu$m, $y_0= 120$ $\mu$m, $\delta=\Gamma/2$. }
%\label{fig:axiTrapVoltagePlot}
%\end{figure}

\section{\label{sec:SBC} Experimental Results of Doppler and Sideband cooling}
\label{sec:expResults}
\subsection{Experimental Setup}
The experiments performed in this paper are realized on a single $^{40}$Ca$^+$ ion confined in a vertical 1.89 T axial magnetic field in a cylindrical, stacked electrode Penning trap (described in more detail in  \cite{Mavadia2013}). Voltages applied to the endcap electrodes provide axial confinement, while the central ring electrode is held at DC ground. This ring is split into four segments to allow for applying AC voltages in order to generate the axialization field used for coupling the radial modes of motion. To apply the axialization potential to the electrodes, a single frequency at $\nu_c=729$ kHz is generated and split into four paths, where the appropriate phase shifts are applied using variable-gain op-amp circuits. 
\par 
Doppler cooling is performed using two diode lasers at 397 nm with vertical polarization on the S$_{1/2,-1/2}\leftrightarrow$ P$_{1/2,+1/2}$ and S$_{1/2,+1/2}\leftrightarrow$ P$_{1/2,-1/2}$ transitions due to the $\sim 20$ GHz frequency splitting of the S$_{1/2}\leftrightarrow$ P$_{1/2}$ transitions. The two lasers have independent frequency control, but are typically set to have the same detuning from resonance. The combined light is split into two paths, one parallel to the magnetic field to cool the axial motion and one perpendicular to the magnetic field to cool the radial motion. The two paths have independent intensity control, but are restricted to having the same frequency. The radial beam position relative to the ion is controlled using a piezo controlled mirror mount with a CCD camera to track the position. The waist of the radial beam is approximately  $w_0\approx100$ $\mu$m. Repumping beams at 866 nm and 854 nm are passed through a fibre electro-optical modulator to generate the necessary frequencies to clear out all Zeeman sublevels of both the D$_{3/2}$ state, and also the D$_{5/2}$ state which is populated due to weakly allowed decay from the P$_{1/2}$ state at high magnetic field \cite{Crick2010}. 
\par 
Coherent addressing is performed on the S$_{1/2,-1/2}\leftrightarrow$ D$_{5/2,-3/2}$ transition with a narrow linewidth ($<1$ kHz) laser at 729 nm with horizontal polarization. The laser frequency is controlled by means of a single-pass acousto-optic modulator driven by a single Direct Digital Synthesizer (DDS) which can dynamically switch between up to 8 frequencies during an experimental cycle, allowing for sideband cooling sequences involving multiple higher order sidebands. The 854 nm laser is also used during the sideband cooling process to quench the D$_{5/2}$ state lifetime and thus increase the cooling rate. 
\par
State preparation is accomplished by optically pumping into the S$_{1/2,-1/2}$ state using the S$_{1/2,+1/2}\leftrightarrow$ P$_{1/2,-1/2}$ laser. State detection uses the standard electron shelving technique, with fluorescence collected on the 397 nm transition using a pair of photomultiplier tubes with typical detection times of 12 ms. Decays from the P$_{1/2}$ state to the D$_{5/2}$ state induce approximately 4\% of shelving background. Each experimental cycle of cooling, probe and detection is repeated 100-200 times at a single frequency.  
\subsection{Experimental Results}

\subsubsection{Doppler cooling}
\begin{figure*}[t]
\centering
\begin{subfigure}[b]{0.45\textwidth}
\centering
\includegraphics[width=1\linewidth]{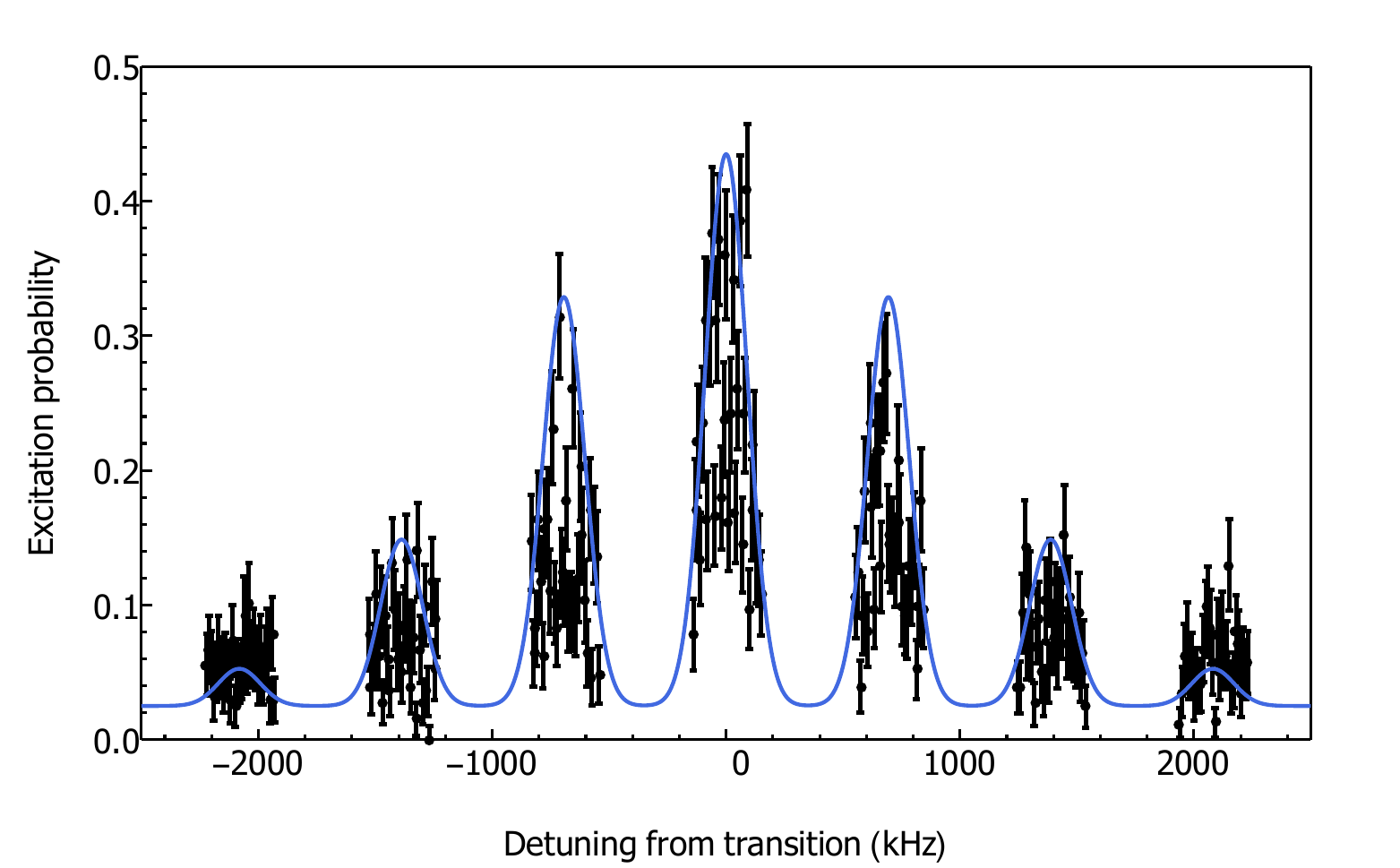}
\caption{Axialization off}
\end{subfigure}
\centering
\begin{subfigure}[b]{0.45\textwidth}
\centering 
\includegraphics[width=1\linewidth]{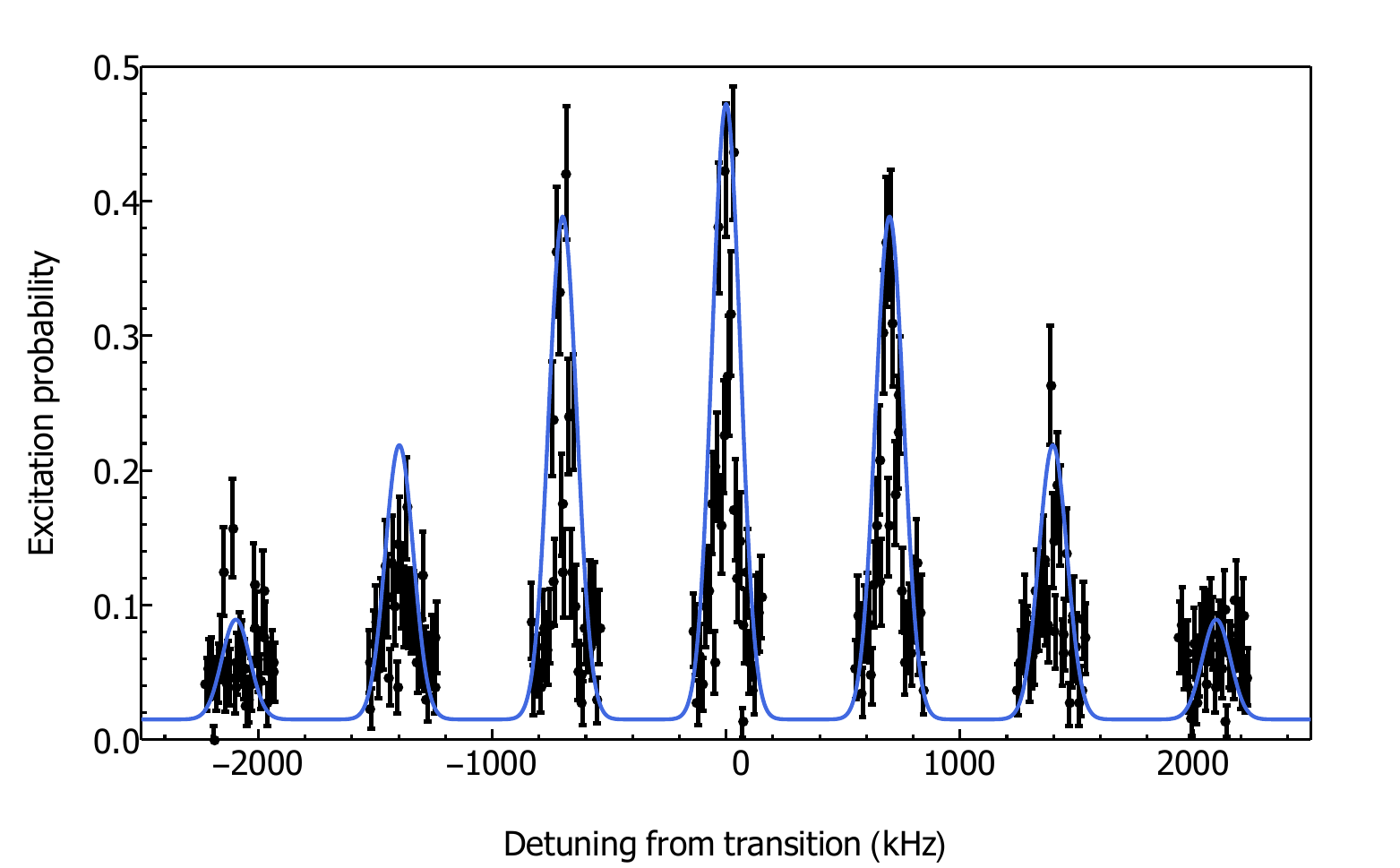}
\caption{Axialization on - 500 mV}
\end{subfigure}
\centering
\begin{subfigure}[b]{0.45\textwidth}
\includegraphics[width=1\linewidth]{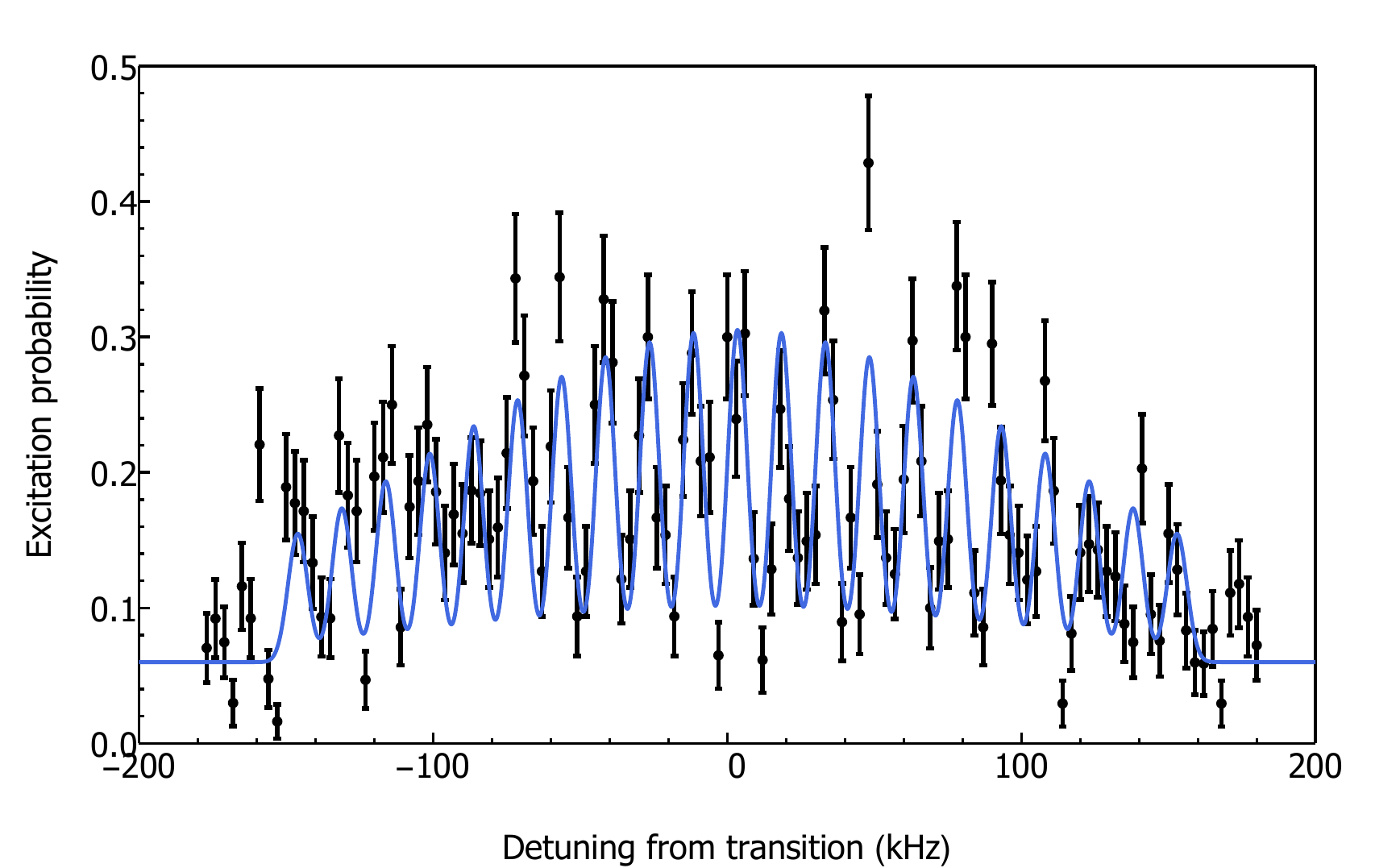}
\caption{Axialization off}
\end{subfigure}
\centering
\begin{subfigure}[b]{0.45\textwidth}
\centering 
\includegraphics[width=1\linewidth]{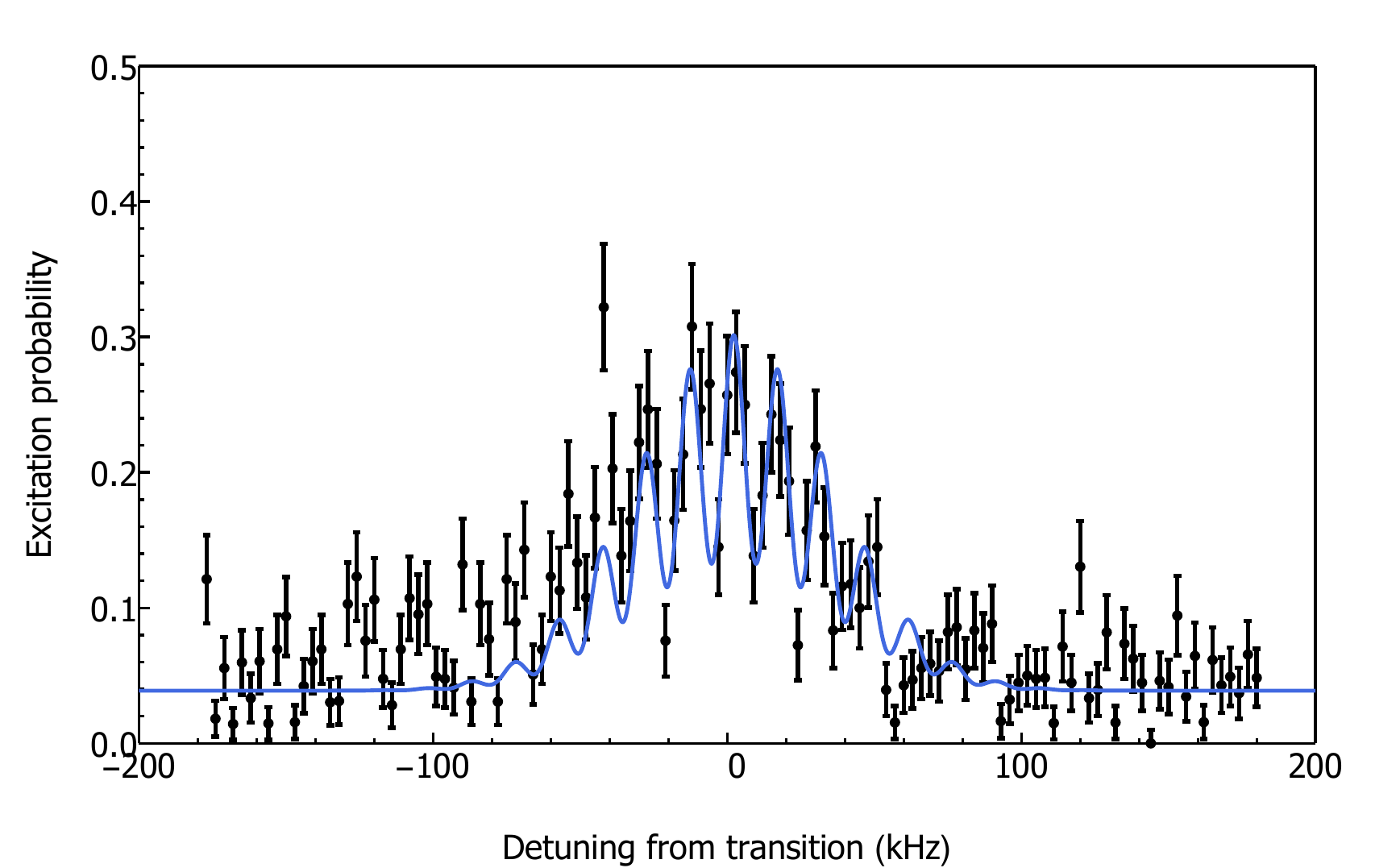}
\caption{Axialization on - 500 mV}
\end{subfigure}
\caption[Radial Spectra of a single ion after Doppler cooling with axialization]{Radial Spectra of a single ion after Doppler cooling at  $\nu_+=693$ kHz, $\nu_-=14.8$ kHz. The average occupation obtained from matching the data to a comb of Gaussian sidebands modulated by a Gaussian envelope: from (a) and (b) $\bar{n}_{+}\approx31$ and $\bar{n}_{+}\approx45$, and  from (c) and (d) $\bar{n}_{-}\approx900$ and $\bar{n}_{-}\approx100$ respectively.  All other parameters are kept same between figures. 500 mV of axialization RF is applied in (b) and (d)}
\label{fig:axiOnOff}
 \end{figure*} 
In order to extract a temperature of the ion after Doppler cooling, we use a fixed length probe and scan the 729 nm laser frequency to obtain excitation strengths of the spectral components of the two radial modes of motion. This approach permits us to decouple the contributions of the two different modes of motion, which would not be possible using Rabi oscillations on the carrier. In the interest of comparing the performance of cooling with and without axialization we must work in a regime where offset beam Doppler cooling still yields a stable and measurable $n_-$ value, which restricts us to working with very low magnetron frequencies on the order of $\nu_z=5-15$ kHz. However, due to the density of spectral features, a fit to the full spectral lineshape summed over the motional Fock states of motion becomes unfeasible.
%of both modes from separate scans using different frequency sampling. and compare the
Instead, we extract the temperatures of both modes by measuring the spectral width of the sideband structure of each mode from separate scans using different frequency sampling. The data is modelled as a comb of Gaussian peaks modulated by a Gaussian envelope with linewidth $\sigma_{\pm}$, which defines the temperature as
\begin{equation}
    T_{\pm}=\frac{M\lambda_{729}^2\sigma_{\pm}^2}{k_B},
\end{equation}
where $\lambda_{729}$ is the transition wavelength. The mean occupation numbers can then be obtained:
\begin{equation}
      \bar{n}_{\pm}=\frac{T_{\pm}k_B\omega_1 }{2\hbar\omega_{\pm}^2}.
\end{equation}

\par
Using the  approach  described above, figure \ref{fig:axiOnOff} compares the mean phonon numbers after 12 ms of Doppler cooling with and without axialization. Firstly, panel (a) exhibits modified cyclotron sidebands spaced by 693 kHz, which are modulated by 15 kHz magnetron sidebands that are not fully resolved. The estimated Doppler temperature of the modified cyclotron motion is within a factor 2 of the Doppler limit. When we zoom in on the carrier, shown in panel (c), we see at least 10 magnetron sidebands, which correspond to a magnetron mean phonon number of approximately $\bar{n}_-\approx 900$.  Applying an axialization drive of $V_{ax}=0.5$ V greatly suppresses the number of magnetron sidebands in panel (d) reducing $\bar{n}_-$ by approximately a factor of 9, while the modified cyclotron spectrum in panel (b) sees only a slight broadening. Thus the net effect of the axialization drive is the removal of over 800 phonons from the total radial motion while bringing the two phonon numbers close to equality.

\begin{figure*}[t]
\centering
\includegraphics[width=.9\linewidth]{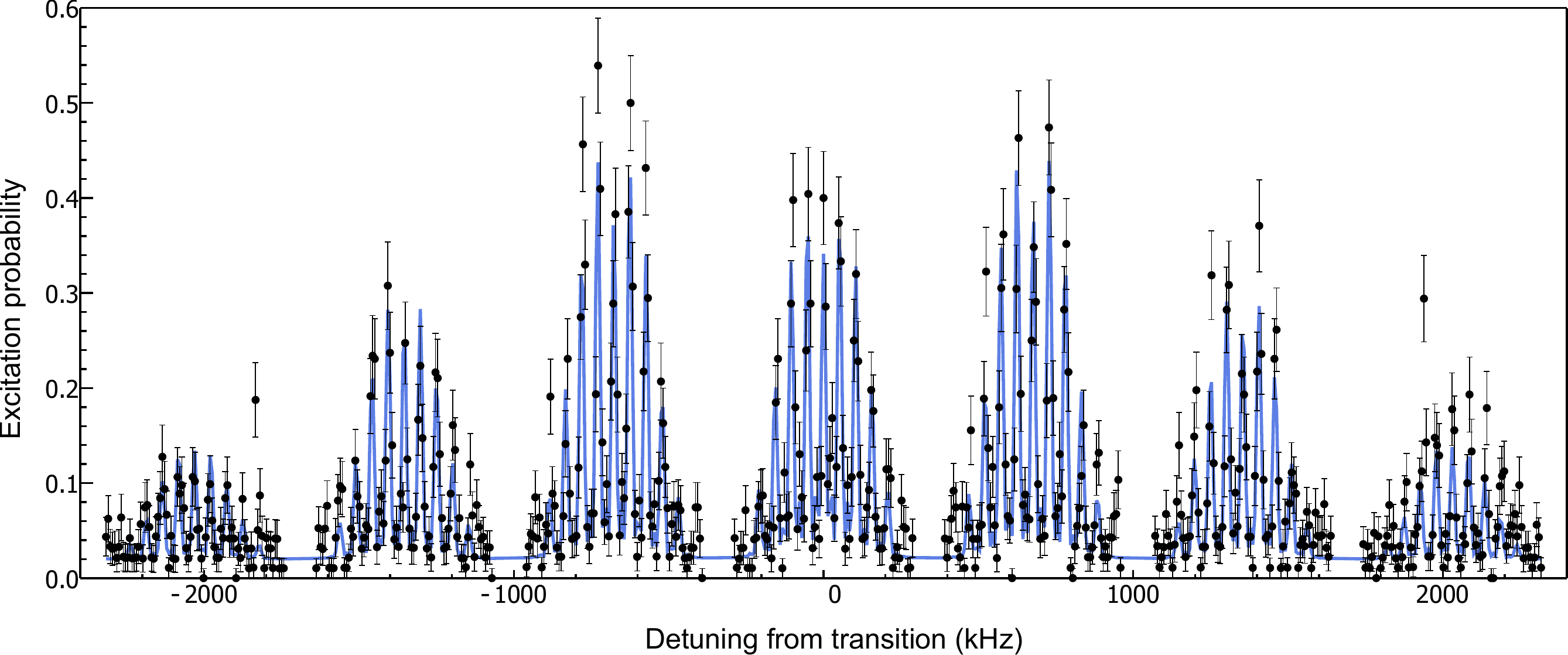} 
\caption[Fitted radial Doppler spectrum at 100V]{Radial Doppler spectrum with $\nu_+=677$ kHz, $\nu_-=52$ kHz taken with 1 V axialization. Full fit to two-mode Rabi dynamics gives the mean phonon numbers $\bar{n}_+=96(5)$ and $\bar{n}_-=136(8)$ and  $\Omega_0/2\pi=26$ kHz }
\label{fig:Doppler100Vrad}
\end{figure*}
\par 
To further confirm that axialization can be used to effectively cool at large trapping frequencies where resolved sideband cooling can be effectively employed, a data set at $\nu_-=52$ kHz and $V_{ax}=1$ V is shown in figure \ref{fig:Doppler100Vrad}. At this frequency, cooling with just an offset beam is not possible. By fitting the full Rabi lineshape dynamics summed over a thermal distribution of phonons to 3 modified cyclotron and 4 magnetron sidebands, mean phonon numbers of $\bar{n}_+=96(5)$ and $\bar{n}_-=136(8)$ were extracted. Once again, the mean phonon numbers approach equality, with the modified cyclotron being slightly lower, as predicted in figure \ref{fig:axvsVolt}, 
%and \ref{fig:axiTrapVoltagePlot},
however they are approximately a factor of 3 larger than the simulated values. One contribution to the discrepancy is that heating due to spontaneous emission from the axial cooling laser was omitted from the simulation. Based on the analytical model of modified cyclotron Doppler cooling this effect could contribute up to 30\% higher mean phonon numbers given a 1:1 scattering ratio of the radial and axial beams. The remaining discrepancy could be attributed to the uncertainty in setting the optimal laser parameters including detuning, beam waist and offset. Nonetheless, axialization allows us to cool the ion to low motional phonon numbers with high reliability and robustness against perturbations such as background gas collisions, with only a modest applied voltage.
%that does not induce appreciable micromotion, as evidenced by any lack of distinctive excitation at the drive frequency of 729 kHz or multiples thereof. 
\subsubsection{Sideband cooling}
As the previous section demonstrated, with the aid of axialization we can consistently reach a regime with $\bar{n}\sim 100$ phonons in both modes at a magnetron frequency  $\nu_-=52$ kHz, which is far outside the Lamb-Dicke regime ($\eta_r\sqrt{2n+1}\approx2$, with $\eta_r$ being the radial Lamb-Dicke parameter). If we take the mean phonon number of $\bar{n}_-=136(8)$ from figure \ref{fig:Doppler100Vrad}, then we find that almost 25\% of the thermal population is in Fock states above the first order sideband coupling minimum at $n=196$, thus clearly ruling out cooling on just first-order motional sidebands. Sideband cooling from outside the Lamb-Dicke regime is still possible, but requires the addressing of higher order motional sidebands to prevent any accumulation of population in any of the coupling minima. This has been shown previously in RF traps \cite{Morigi1999,Poulsen2012b} as well as in our Penning trap for the axial motion of single ions \cite{Goodwin2016} and small Coulomb crystals \cite{Stutter2018}. In fact, cooling the axial modes of a two-ion chain at low trap frequency is similar to the problem of cooling the two radial modes of a single ion in that both require addressing higher order sidebands of both motions in turn. However, in this case we need to address the blue sidebands of the magnetron motion, rather than the red sidebands, as a consequence of the negative total energy of this motion. Following the approach to cooling sequence design for two modes of an ion chain outlined in reference \cite{Joshi2019}, we adopt a 68 ms long cooling sequence (Table \ref{table:radCoolTab}) that addresses the red modified cyclotron and blue magnetron sidebands. This sequence is not proven to be optimal, but is tested to be experimentally robust and follows the philosophy that cooling is initially faster on higher order sidebands, while the lowest order sidebands are required to reach the lowest mean phonon number. Furthermore, owing to the proximity of the first magnetron sideband to the carrier, the 729 nm laser Rabi frequency is lowered by approximately a factor of $\sqrt{2}$ and the effective upper state linewidth set by the quench laser is maintained at $\Tilde{\Gamma}/2\pi=5-10$ kHz to minimize unwanted off-resonant excitation of the carrier. Figure \ref{fig:Doppler100Vrad} suggests that sidebands that reduce the motional quanta of both modes could be initially used to cool both modes simultaneously when far outside the Lamb-Dicke regime, but this approach was not pursued here. 
\par
\begin{figure}[h]
\centering
\begin{subfigure}[b]{0.45\textwidth}
\centering
\includegraphics[width=1\linewidth]{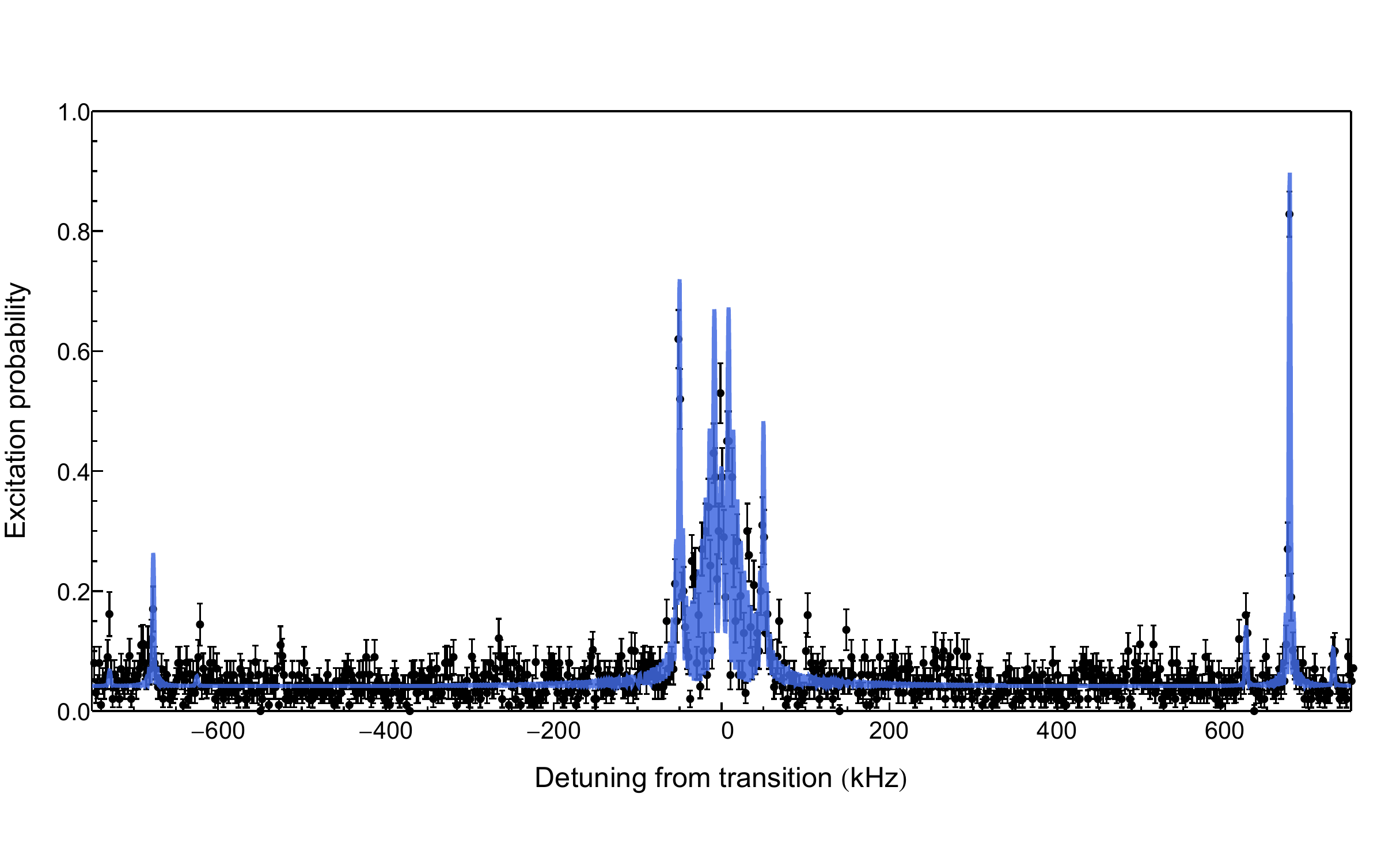}
\end{subfigure}
\centering
\begin{subfigure}[b]{0.45\textwidth}
\centering 
\includegraphics[width=1\linewidth]{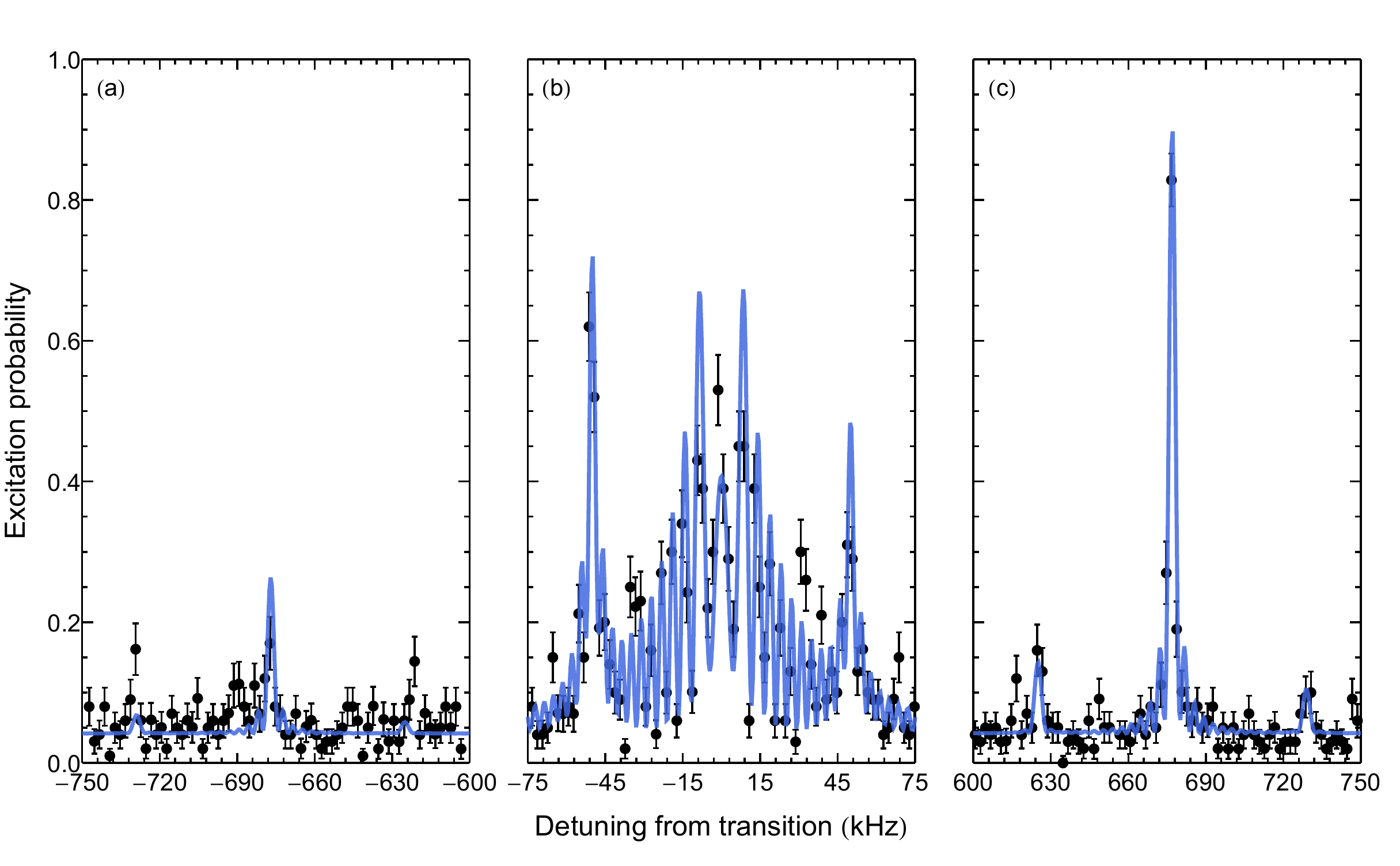}
\end{subfigure}
\caption{\textit{Top} -- Continuous spectrum after 68 ms of sideband cooling (see table \ref{table:radCoolTab}) with a probe time of 280 $\mu$s. \textit{Bottom} -- Zoomed in plots around (a) 1\textsuperscript{st} red cyclotron sideband (b) carrier with magnetron sidebands and (c)  1\textsuperscript{st} blue cyclotron sideband. Fit parameters using a two-mode thermal model: $\bar{n}_+=0.35(5)$, $\bar{n}_-=1.7(2)$, $\Omega_0/2\pi=14.13(4)$ kHz.  }
\label{fig:radSBCfull}
\end{figure}

After applying the sequence of table \ref{table:radCoolTab} to an ion cooled with the same Doppler beam and trap parameter settings as in figure \ref{fig:Doppler100Vrad}, the spectrum shown in figure \ref{fig:radSBCfull} is obtained. The strong suppression of the red modified cyclotron sideband and the asymmetry in the magnetron sidebands is evidence that a significant population of both modes is in the ground state. A fit to the Rabi dynamics summed over a thermal Fock state population results in phonon numbers of $\bar{n}_+=0.30(5)$, $\bar{n}_-=1.7(2)$. Furthermore, the spectrum of figure \ref{fig:radSBCfull} shows no higher order red magnetron sidebands as would be expected if population were trapped at a coupling minimum. The red modified cyclotron sideband does appear to have two excess shelving events near $\nu_+\pm\nu_-$, however they are not consistent with a thermal model fit, nor can they represent trapped population in the magnetron coupling minima as there is no evidence of corresponding magnetron sidebands around the carrier. An additional scan was later performed around the modified cyclotron sidebands where these events were not reproduced. Since the cooling sequence ends with a modified cyclotron cooling pulse, the $\bar{n}_-$ also has an appreciable contribution due to motional heating, which was separately measured to be $\approx 300$ phonons s$^{-1}$. The motional heating is most likely to originate from technical noise on the DC supplies of the ring electrodes. The remaining excess thermal population of the magnetron mode arises from off-resonant carrier excitation and could be further suppressed by working at a higher magnetron frequency.  
\begin{table}
\begin{center}
\begin{tabular*}{0.4\textwidth}{c|c|c}
\hline 
Sideband & Pulse length (ms) & 729 Intensity (\%) \\ 
\hline
2\textsuperscript{nd} Cyc. & 5 & 100\\
1\textsuperscript{st} Cyc. & 10 & 100 \\
\hline
3\textsuperscript{rd} Mag. & 5 & 100 \\
2\textsuperscript{nd} Mag. & 5 & 100 \\
1\textsuperscript{st} Mag. & 5 & 52 \\
\hline
2\textsuperscript{nd} Cyc. & 5 & 100\\
1\textsuperscript{st} Cyc. & 5 & 100 \\
\hline
3\textsuperscript{rd} Mag. & 5 & 100 \\
2\textsuperscript{nd} Mag. & 5 & 100 \\
1\textsuperscript{st} Mag. & 5 & 52 \\
\hline
1\textsuperscript{st} Cyc. & 10 & 100\\
1\textsuperscript{st} Mag. & 2 & 52 \\
1\textsuperscript{st} Cyc. & 1 & 100 \\
\hline 
\end{tabular*} 
\caption{68 ms sideband cooling sequence for both radial modes.}
\label{table:radCoolTab}
\end{center}
\end{table}

\par
\begin{figure}
\centering
\begin{subfigure}[b]{0.45\textwidth}
\centering
\includegraphics[width=1\linewidth]{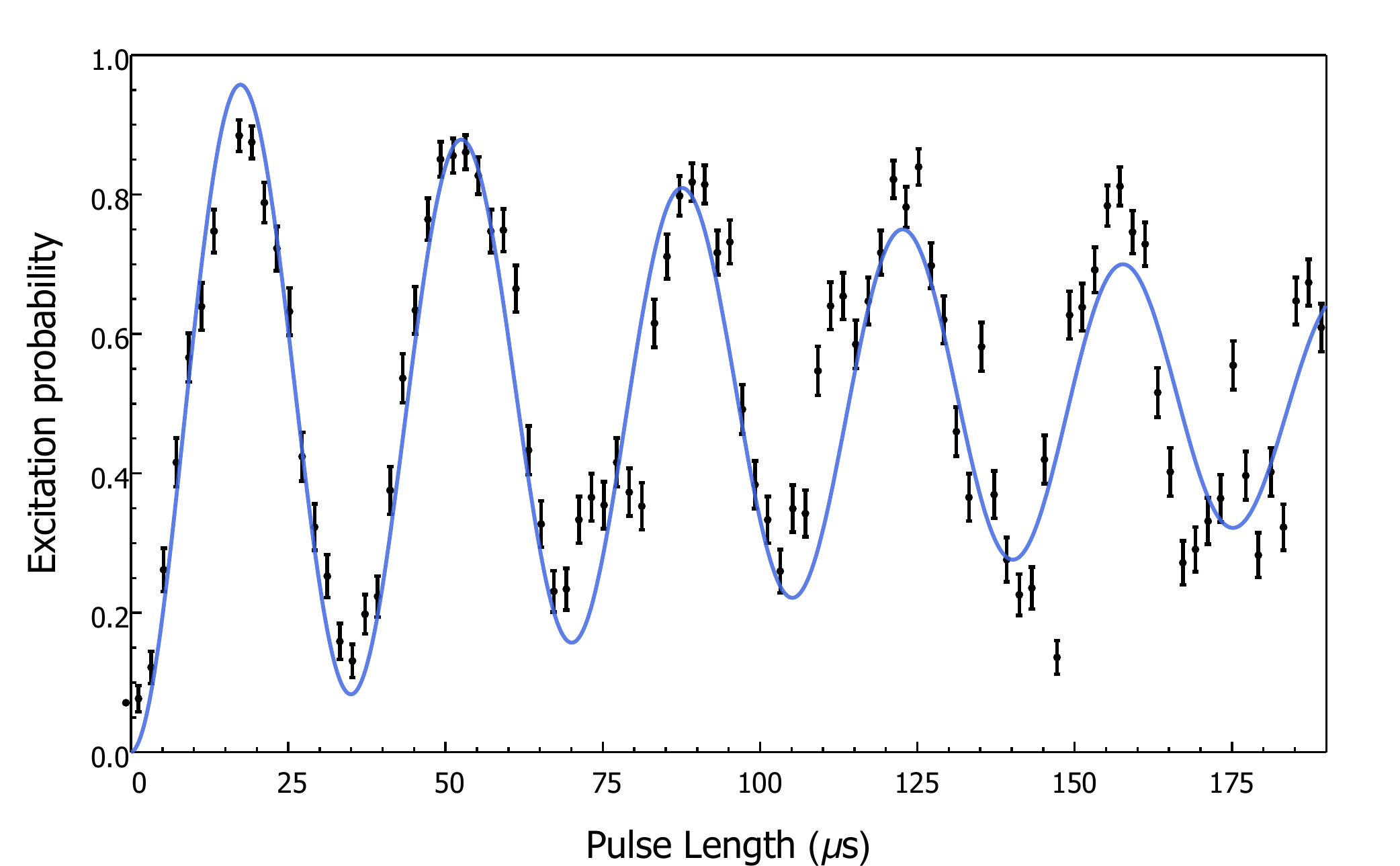}
\end{subfigure}
\centering
\begin{subfigure}[b]{0.45\textwidth}
\centering 
\includegraphics[width=1\linewidth]{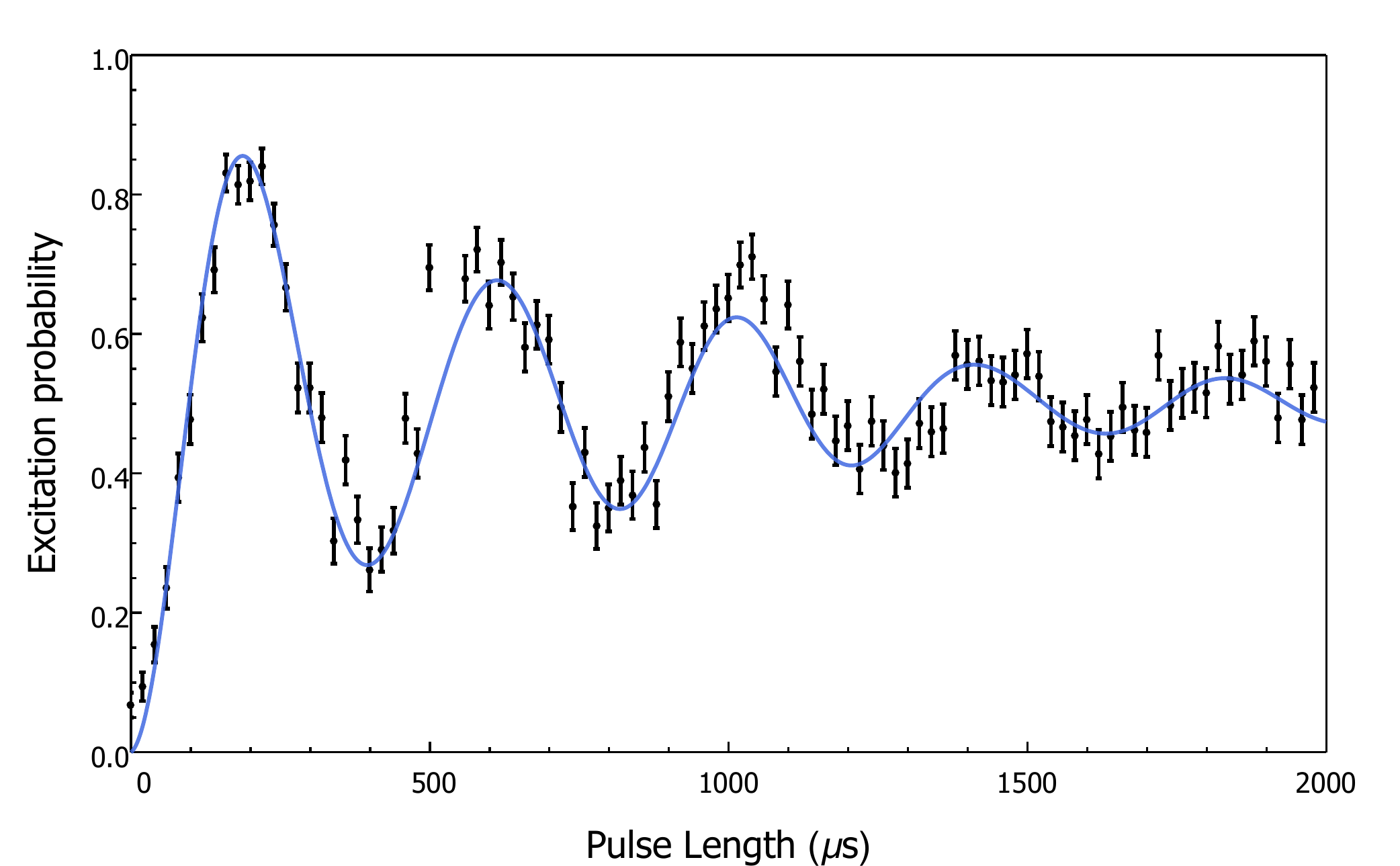}
\end{subfigure}
\caption[Continuous SBC fit -- 68 ms cooling]{Rabi oscillations  after 68 ms of sideband cooling. \textit{Top} -- Carrier.  \textit{Bottom} -- Modified cyclotron blue sideband. Fit parameters using a two-mode thermal model with fixed $\bar{n}_-=1$: $\bar{n}_+=0.15(3)$, $\Omega_0/2\pi=20.70(7)$ kHz.  }
\label{fig:radRabi}
\end{figure}

 
We also verify our ability to coherently drive the ion by performing Rabi oscillations on the carrier and the modified cyclotron blue sideband, as shown in figure \ref{fig:radRabi}. The contrast of the carrier oscillations is limited by fast Rabi frequency fluctuations of $3.1\%$ extracted from an exponential decay of the visibility, with slow noise also causing measurement points at later times to lie several standard deviations away from the fit line. The source of this effect is believed to be polarization noise on the final 5 meter long optical fibre of the spectroscopy laser before the trap where the transition is sensitive to any drift away from perfect linear polarization. Due to this noise, the carrier Rabi oscillation data cannot be used to accurately constrain the mean phonon numbers of the two modes as Rabi frequency components only depend weakly on $n$. On the other hand, the modified cyclotron Rabi oscillations will have frequency components that depend strongly on $\bar{n}_+$ and the decay lineshape for a thermal distribution of $\bar{n}_+\approx 1$ can be distinguished from an exponential decay induced by fast Rabi frequency noise. Using a fit to the two-mode Rabi excitation formula: 
\begin{equation}
\begin{split}
P_e(t)=\sum_{n_+,n_-}^{15}\frac{\bar{n}_+^{n_+}}{(\bar{n}_++1)^{n_++1}}\frac{\bar{n}_-^{n_-}}{(\bar{n}_-+1)^{n_-+1}}\times
\\
\frac{1}{2}\left(1-e^{-t/\tau_e}\cos\Omega_{n_+,n_+}^{n_-,n_-} t\right)
\label{eq:radialRabi}
\end{split} 
\end{equation}
with the generalized Rabi frequency $\Omega_{n_+,n_++1}^{n_-,n_-}$ phonon dependence incorporated using the full Laguerre polynomial dependence \cite{Wineland1998} and an exponential decay time constant $\tau_e$, the data suggests $\bar{n}_+=0.15(3)$ for $\bar{n}_-$ fixed to 1. The decay constant of $\tau_e=780(40)$ $\mu$s corresponds to a generalized Rabi frequency noise of $10\%$, which is consistent with the polarization fluctuations described above as well as additional broadening due to the laser linewidth of $\approx 1$ kHz. 
\section{\label{sec:conclusion}Conclusion}

In this article we have presented experimental results of sideband cooling showing significant occupation of the ground states of motion of both radial modes of a single ion in a Penning trap. Due to the negative energy of the magnetron motion,  blue motional sidebands were excited in order to remove motional quanta. On the other hand, the modified cyclotron was cooled in the conventional manner where red sidebands of motion were used. As a pre-requisite for this result, we experimentally show how the use of an axialization drive providing a resonant mode coupling works together with a radially offset Doppler cooling beam to effectively reduce the mean phonon number of the magnetron motion to values amenable to sideband cooling.  Furthermore, we numerically study the parameter space of Doppler cooling and show that with modest applied axialization voltage, low mean phonon numbers of both modes can be attained. Coherent control of a ground state cooled ion is demonstrated by showing Rabi flops on the carrier and blue sideband transitions. The studies are useful for future Penning trap based high precision and quantum simulations experiments, where the unwanted effects due to occupation in motional states can be minimized by performing sideband cooling.
\section*{\label{sec:Ack}Acknowledgements}
The research leading to these results has received funding  from  the  People  Programme  (Marie  Curie  Actions) of the European Union’s Seventh Framework Programme  (FP7/2007-2013)  under  REA  grant  agreement no.  31723. This  work  was  supported  by  the  UK  Engineering and Physical Sciences Research Council (Grant EP/L016524/1)
%\begingroup
%\renewcommand{\arraystretch}{1.0}

%\endgroup
% \section{\label{sec:appendix}Appendix}
% \begin{figure}[H]
% \centering
% \includegraphics[width=.9\linewidth]{radialSBCwindowLow.pdf} 
% \caption[Windowed modified cyclotron SBC fit -- 66 ms cooling]{Spectrum after 66 ms of sideband cooling  probed for 280 $\mu$s. Fit parameters to the two mode thermal model: $\Omega_0/2\pi=29.50(11)$ kHz, $\bar{n}_+=0.28(6)$, $\bar{n}_-=1.9(4)$. }
% \label{fig:radialSBCwindowSBonly}
% \end{figure}
\input{BBLAPSpaper.bbl}
\end{document}